%% file: arxiv_build.tex
\documentclass[aps,prb,onecolumn,floatfix,superscriptaddress]{revtex4-2} 
\usepackage{graphicx}
\usepackage{subfigure}
\usepackage{mathtools}
\usepackage{booktabs}
\usepackage{ulem}
\usepackage[T1]{fontenc}
\usepackage[utf8]{inputenc} 
\usepackage{fancyhdr}
\usepackage[export]{adjustbox}
\usepackage{dsfont}
\usepackage{setspace}
\usepackage[colorlinks=true,linkcolor=black,urlcolor=black,citecolor=black]{hyperref}
\usepackage{makeidx}
\usepackage[toc]{appendix}
\usepackage{bm}
\usepackage{multirow}
\usepackage{xr}

\usepackage{changes}

\newcommand{\blangle}{\bigl\langle}
\newcommand{\brangle}{\bigr\rangle}
\newcommand{\dlangle}{\langle\kern-1.5pt\langle}
\newcommand{\drangle}{\rangle\kern-1.5pt\rangle}
\newcommand{\bdlangle}{\blangle\kern-3pt\blangle}
\newcommand{\bdrangle}{\brangle\kern-3pt\brangle}

\begin{document}

\input{arxiv_multifractality}        
\clearpage
\appendix
\input{arxiv_supplementary}  
\bibliography{biblio_multifractality}
\end{document}

%% file: arxiv_multifractality.tex
\title{Emergence of non-ergodic multifractal quantum states in geometrical fractals}
\author{Fabio Salvati}
\email{fabio.salvati@ru.nl}
\affiliation{Institute for Molecules and Materials, Radboud University, Heijendaalseweg 135, 6525 AJ Nijmegen, The Netherlands}

\author{Mikhail I. Katsnelson}
\affiliation{Institute for Molecules and Materials, Radboud University, Heijendaalseweg 135, 6525 AJ Nijmegen, The Netherlands}
\affiliation{Constructor Knowledge Institute, Constructor University, Campus Ring 1, 28759 Bremen, Germany}

\author{Andrey A. Bagrov}
\affiliation{Institute for Molecules and Materials, Radboud University, Heijendaalseweg 135, 6525 AJ Nijmegen, The Netherlands}

\begin{abstract}
    Eigenstate multifractality, a hallmark of non-interacting disordered metals, which may also be observed in many-body localized states, is characterized by anomalous slow dynamics and appears relevant for many areas of quantum physics, from measurement-driven systems to superconductivity. We propose a novel approach to achieve non-ergodic multifractal states (NEMs) without disorder by iteratively introducing defects into a crystal lattice, reshaping it from a plain structure into a fractal geometry. By comprehensive analysis of the Sierpiński gasket case, we find robust evidence of the emergence of NEMs that go beyond the conventional classification of quantum states and designate new pathways for quantum transport studies. We discuss potential experimental signatures of these states. 
\end{abstract}

\maketitle

\section*{Introduction} 

Since the discovery of Anderson's localization paradigm \cite{localization_anderson}, there has been a significant interest in critical phenomena that transcend the traditional dichotomy of extended and localized states. Among these, a class of novel states, commonly referred to as non-ergodic multifractal states (NEMs) \cite{localization_mirlin, localization_peliti}, has attracted particular attention. NEMs occupy a finite volume of the system, yet their support set, in the thermodynamic limit, has zero measure, challenging the conventional ideas of how a quantum state can be distributed in space. The concept of multifractality finds broad applications across science and engineering, including fluid dynamics \cite{fluids_frisch}, topography \cite{topography_gerges}, finance \cite{finance_schmitt}, physiology \cite{physiology_ivanov}, and epidemic modeling \cite{epidemy_long}. It also inspired advancements in machine learning, granting unexplored settings to incorporate greater hierarchical depth and achieve stable prediction performances \cite{machinelearning_wang, machinelearning_yu}. In condensed matter physics, the concept of NEMs is strictly connected to the singular continuous nature of the spectrum \cite{singspec_simon, singspec_jitomirskaya, singspec_kravtsov} and the critical scaling behavior \cite{localization_mirlin, sizescaling_rodriguez} observed at phase transitions, which give rise to anomalous subdiffusive dynamics \cite{transport_luitz, multifractality_cugliandolo} and make them particularly intriguing in the context of quantum transport. Additionally, they exhibit sublinear growth of entanglement entropy \cite{mbl_agarwal, multifractality_detomasi}, defying the standard formulation of the eigenstate thermalization hypothesis \cite{mbl_orito}. These attributes are essential for understanding quantum coherence of eigenstates \cite{multifractality_berthold} and many-body localization phenomena \cite{mbl_mace, mbl_pietracaprina}. Moreover, it is conjectured that multifractality can enhance electron-electron interactions, leading to elevated critical temperature in superconductors \cite{superconductivity_feigelman, superconductivity_rubioverdu}.

Incommensurate potentials in one-dimensional systems \cite{localization_aubry, quasiperiodic_bloch} and random disorder in higher-dimensional systems \cite{localization_thouless, localization_deluca, localization_zirnbauer} are established mechanisms for inducing NEMs. Experimental investigation of these routes, however, often faces notable hurdles. For instance, with incommensurate systems, achieving ideal mathematical conditions is difficult. Practical realizations frequently employ rational approximations whose periodicity can introduce finite-size effects or mini-gaps, potentially masking true multifractal signatures \cite{quasiperiodic_sokoloff, quasiperiodic_bloch}. Systems relying on random potentials, on the other hand, demand precise critical tuning of disorder \cite{disorder_kropf}. Moreover, observations from single experimental instances may not capture the true ensemble-averaged multifractal characteristics, leading to considerable sample-to-sample variability. Beyond these specific issues, a more general challenge is that the delicate quantum interference essential for NEMs is highly vulnerable: it can be readily destroyed by slight deviations from ideal parameters, interparticle interactions, or other uncontrolled perturbations. To address these limitations, we propose an alternative approach: generating multifractal states through precise geometric manipulation, by arranging substitutional point-like defects to transform a regular lattice into a deterministic fractal geometry \cite{fractals_mandelbrot}.
In recent years, properties of artificial quantum fractal structures have been systematically explored experimentally and theoretically \cite{fractals_biesenthal, fractals_kempkes, fractals_fischer, fractals_mondal, fractals_canyellas, fractals_hall_iliasov, fractals_ivaki, fractals_yao, fractals_westerhout, fractals_neupert, fractals_supercond_iliasov, fractals_cond_vanveen, fractals_anyons_manna, fractals_topological_manna, fractals_lage}. Here, our aim is to demonstrate how the self-similar structure and non-integer dimensionality of fractals offer a natural framework for designing and studying multifractal eigenstates.

We consider single-particle quantum mechanics of a tight-binding model on a Sierpiński gasket (SG), and through a comprehensive numerical analysis, we demonstrate multiple crossovers between extended states and distinct stages of multifractality in the model spectrum. By extrapolating the fractal dimensions, we identify energy intervals where transitions to multifractal behavior occur. Examination of the singularity spectrum confirms the multifractal nature of these individual states and unveils a variety of behaviors, including NEMs that may retain metallic or insulating features. Recent advancements in spin-resolved scanning tunneling microscopy (STM) and spectroscopy techniques \cite{stm_khajetoorians, stm_sierda} have enabled precise control over quantum states in low-dimensional structures with strong quantum confinement. With this in mind, we also present a detailed numerical analysis of the density-density correlation functions, which can be directly observed in such experiments. By pinpointing the regions where crossovers to NEMs happen, we compute these correlators and show that they exhibit the predicted power-law decay. Altogether, these attributes form a robust criterion to unambiguously distinguish NEMs from both extended and localized states. 

While this study is purely theoretical, one aims to exploit the unconventional properties of NEMs in  contexts like quantum control or superconductivity. Hence, a key question is: can multifractal states generated with geometrical fractals demonstrate sufficient resilience to perturbations to support their practical use? To address this, we introduce additional randomly distributed geometric defects into the system. Our analysis shows that, qualitatively, NEMs exhibit robustness against such perturbations. What is even more compelling is that some NEMs react to manipulation of individual atoms in a very structured and symmetric way, possibly opening an opportunity to use them for delicate control in quantum systems.

\begin{figure*}[ht!]
    \includegraphics[width=0.7\linewidth]{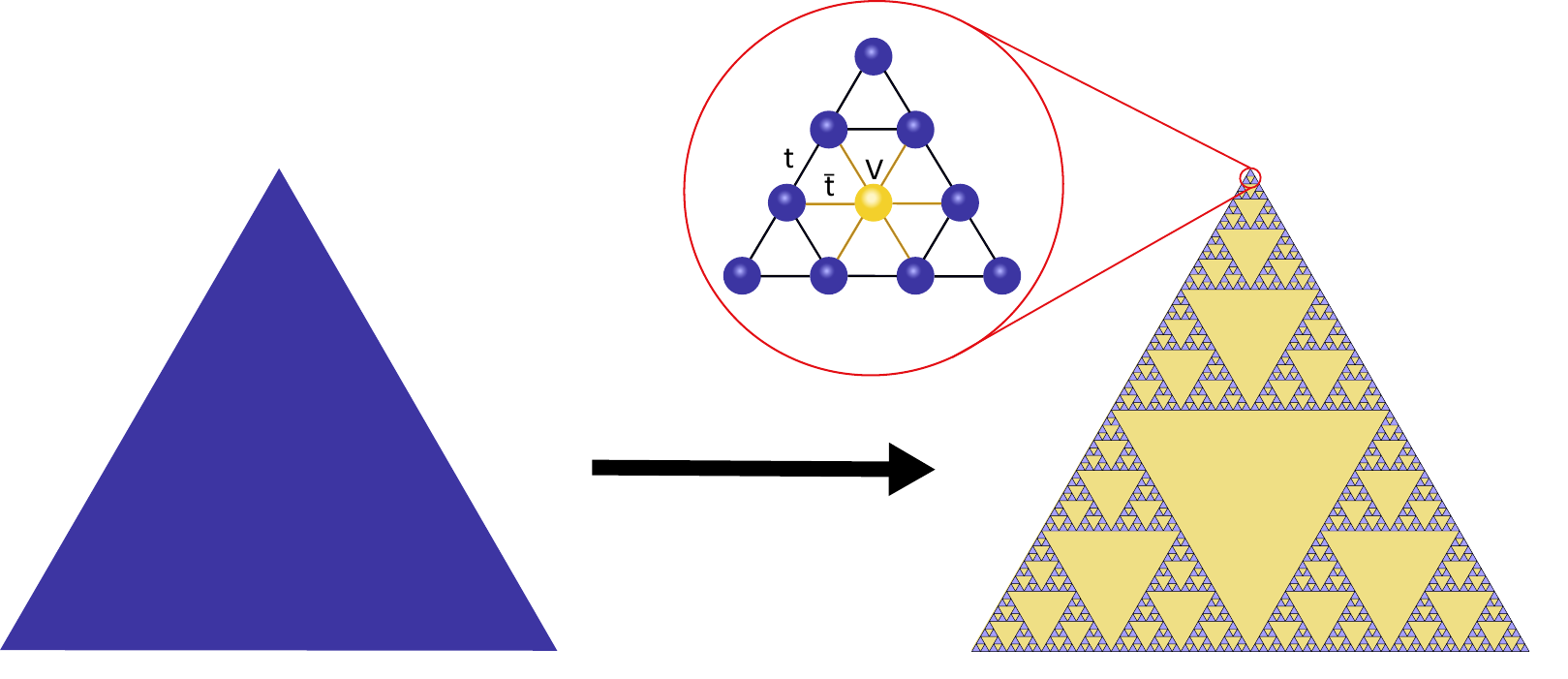}
    \caption{Transformation of regular geometry into a fractal structure: from triangular lattice (left) into a sixth generation Sierpiński gasket (SG, right). Zoom-in: the elementary building block of the SG with a substitutional defect (yellow dot). The hopping between regular sites is $t$. The defect has on-site potential $V$, and the corresponding hoppings to the rest of the lattice are $\bar{t} \ll t \ll V$.}
    \label{fig:triangle_to_gasket}
\end{figure*}

\section*{Model}

Our starting point is a uniform triangular lattice $\mathcal{L}$ containing $N_\mathcal{L}$ sites, each uniquely indexed. The corresponding tight-binding Hamiltonian is an $N_\mathcal{L} \times N_\mathcal{L}$ matrix. With uniform geometry and assuming hopping exclusively between nearest neighbors, off-diagonal Hamiltonian elements are $t_{ij} = t = 1$ for neighboring sites $i$ and $j$ and zero otherwise. The diagonal elements are initially set to zero, neglecting any on-site potential at this stage. The effects of including next-nearest-neighbor (NNN) hopping are discussed in the Supplemental Material (SM, Sec.~\ref{sec:sm_nnn_hopping}). 

The transformation of the triangular lattice into a SG is achieved by iterative placement of defects into regions corresponding to voids in the fractal structure up to the desired generation $g$, as shown in Fig.~\ref{fig:triangle_to_gasket}. The set of defect sites forms a subset of the original lattice denoted as $\mathcal{B} \subset \mathcal{L}$, containing $ N_\mathcal{B}$ sites. The remaining $N_g = N_\mathcal{L} -  N_\mathcal{B}$ sites constitute the active sites of the $g$-th generation SG. Details on the site counts ($ N_\mathcal{L}$, $N_\mathcal{B}$, $N_g$) for the generations studied are provided in the SM, Sec.~\ref{sec:sm_geometry}.
 
Each defect, located at position $r_i$, is modeled as a strong, localized potential: $V_i = V \,\delta(r - r_i)$, with strength $V$ significantly larger than any other energy scales ($V \gg t$). To effectively isolate these defect sites and mimic the voids of the fractal, the hopping terms involving links connected to any defect site are drastically reduced ($\bar{t} \ll t$). The resulting single-particle Hamiltonian reads:

\begin{equation}
    \label{eq:hamiltonian}
    H = \sum\limits_{\substack{\langle i,j \rangle \\ i, j \notin \mathcal{B}}} t\, c^\dagger_i c_j + \sum_{k \in \mathcal{B}} V\, c^\dagger_k c_k + \sum_{\substack{\langle i,k \rangle \\  k \in \mathcal{B} }} \overline{t}\, c^\dagger_i c_k + \text{h.c.}
\end{equation}
    
We compute the eigenvalues and eigenvectors of this Hamiltonian, Eq.~\eqref{eq:hamiltonian}, with open boundary conditions. For systems up to generation $g=7$, the full spectrum is obtained using exact diagonalization. However, this approach becomes infeasible for larger lattices due to the exponential scaling of the Hilbert space dimension. Therefore, for generations $g \ge 8$, we employed Krylov subspace methods, as implemented in the SLEPc library~\cite{slepc}, to compute eigenvalues and eigenvectors within specific energy ranges of interest.

For the results presented here, the model parameters are set to $\bar{t} = 10^{-9}$, $t = 1$, and $V = 10^4$. We emphasize that this choice is representative, as any set of values satisfying the inequality $\bar{t} \ll t \ll V$ is sufficient to reproduce the same qualitative results.
 
\section*{Results}
\subsection*{Eigenstates Statistics}

The degree of delocalization in a quantum system can be probed by examining the structure of its eigenstates $\Phi$. 
A standard approach, derived from multifractal analysis~\cite{localization_mirlin, sizescaling_rodriguez}, is the Inverse Participation Ratio ($\text{IPR}_q$). For a given quantum state, it is defined as: 

\begin{equation}
    \label{eq:ipr_definition}
    \text{IPR}_q(\Phi) = \sum_{i=1}^{N} |\langle i | \Phi \rangle|^{2q}.
\end{equation}

Here, $N$ is the total number of available sites of the SG (i.e. $N = N_g$), and $q$ is a real parameter, referred to as the moment order. While Eq.\eqref{eq:ipr_definition} is basis-dependent, in what follows we will employ the basis of localized states, with $|i\rangle$ wavefunction being fully concentrated at site $i$. 

In the thermodynamic limit ($N \to \infty$), $\text{IPR}_q$ is expected to scale with the system size as:

\begin{equation}
    \label{eq:scaling_relation}
    \text{IPR}_q(\Phi) \sim N^{-\tau_q}.
\end{equation}

The exponents $\tau_q$, known as mass exponents, are related to the generalized fractal dimensions $D_q$ by $\tau_q = D_q(q-1)$. These dimensions describe the spatial extent of the eigenstates and distinguish between different classes of quantum states. Within this framework, the dimensions are normalized to the geometry under study, ensuring that $D_q$ directly quantifies the fraction of sites occupied by the eigenstate~\cite{fractdim_rammal, fractdim_orbach}.
Extended (metallic-like) states spread across the entire lattice; their wave function intensity at any site $i$ typically scales as $|\langle i | \Phi \rangle|^2 \sim 1/N$. This leads to $\text{IPR}_q \sim N \cdot (1/N)^q = N^{1-q}$, which implies $\tau_q = q-1$ and consequently $D_q = 1$ for all $q$. Conversely, localized (insulating) states occupy a finite region, $\mathcal{O}(1)$ sites, independent of $N$. As a result, $\text{IPR}_q$ approaches a constant value in the thermodynamic limit, yielding $\tau_q = 0$ and thus $D_q = 0$ for all $q>0$. Multifractal states represent an intermediate scenario between these two extremes. These states are non-ergodic, populating only a fractal subset of the available sites. They are characterized by mass exponents $0 < \tau_q < q-1$ (or equivalently: $0 < D_q < 1$) and exhibit non-trivial $q$-dependence.

The moment order $q=2$ in Eq.~\eqref{eq:ipr_definition} is a widely used quantity to assess the localization of eigenstates. To facilitate comparisons and visualization, we introduce a normalized version \cite{multifractality_qi}, $\text{nPR}_2(\Phi)$, that reads as:

\begin{equation}
    \label{eq:nPR2_definition}
    \text{nPR}_2(\Phi) = \frac{1}{N \cdot \text{IPR}_2(\Phi)} = \frac{1}{N \sum_{i=1}^N |\langle i | \Phi \rangle|^4}
\end{equation}

\begin{figure}[h!]
    \includegraphics[width=0.8\linewidth]{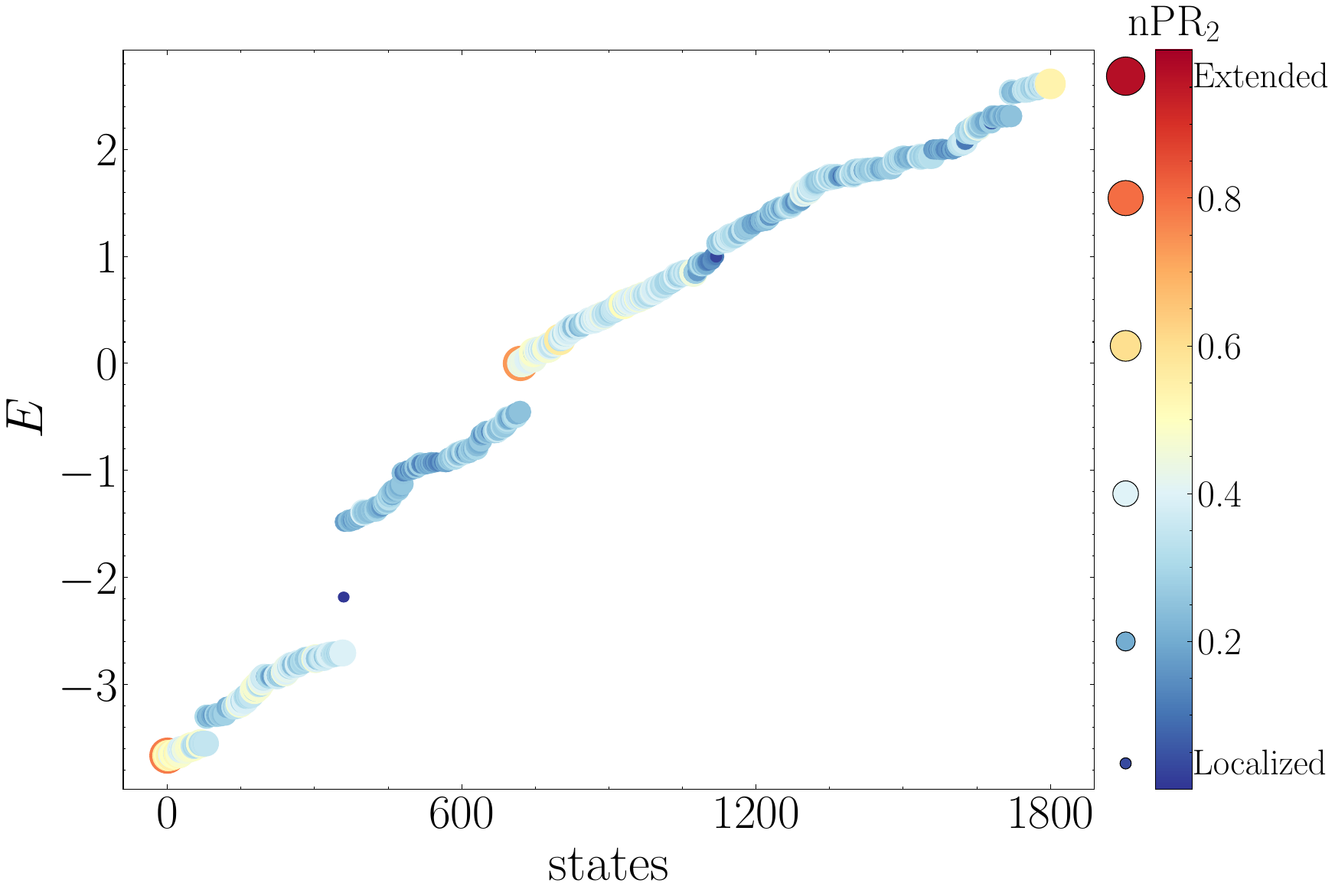}
    \caption{Normalized Participation Ratio ($\text{nPR}_2$) for eigenstates of the sixth-generation Sierpiński gasket. States are plotted by energy $E$ versus state index. The color and marker size denote $\text{nPR}_2$: high values (dark red, large circles) are indicative of extended states, and low values (dark blue, small circles) of localized states. Owing to the spectrum’s hierarchical structure (SM, Sec.~\ref{sec:sm_hierarchy}) distinct regimes emerge, with an abundance of critical states.}
    \label{fig:nPR_gasket}
\end{figure}

In the thermodynamic limit, $\text{nPR}_2$ approaches $1$ for extended states ($\text{IPR}_2 \sim N^{-1}$) and $0$ for localized states ($\text{IPR}_2 \sim \mathcal{O}(1)$). We apply this analysis to the eigenstates of the SG tight-binding Hamiltonian (Eq.~\eqref{eq:hamiltonian}), and Fig.~\ref{fig:nPR_gasket} shows the computed $\text{nPR}_2$ values for all such states from the $g=6$ generation, plotted against their corresponding energies.

For the aforementioned case, the $\text{nPR}_2$ distribution reveals a rich energy-dependent localization landscape. For example, the ground state and several low-energy excited states exhibit high $\text{nPR}_2$ values, indicative of a possible extended nature within this finite system. Localization is suggested for certain energies— at $E\simeq -2.184$ and $E \simeq 1.079$, where the eigenstates have $\text{nPR}_2\simeq 0$.
A considerable fraction of eigenstates possesses intermediate $\text{nPR}_2$ values— spanning the $(0,1)$ interval. These hint at complex, non-uniform spatial distributions that deviate from simple extended or localized forms, a characteristic feature of critical states. Such behavior is particularly anticipated in high spectral density regions or near band gaps, where the fractal’s self-similar geometry shapes eigenstates and supports multifractality~\cite{spectral_dasilva}.

\subsection*{Finite-Size Scaling and Multifractal Analysis}

Based on the $\text{nPR}_2$ overview of the $g=6$ generation (Fig.~\ref{fig:nPR_gasket}), we select three representative energies to carry out a detailed finite-size scaling of the corresponding states, extrapolating their behavior to the thermodynamic limit. These energies are associated with: the ground state (GS), which has a high $\text{nPR}_2$ value; the degenerate triplet at $E\simeq -2.184$, which exhibits the lowest $\text{nPR}_2$ values; and eigenstates at $E\simeq 1.509$, chosen from a spectral region with intermediate $\text{nPR}_2$ values as potential multifractal candidates. Our primary goal is to determine the mass exponents, $\tau_q(E)$, and the generalized fractal dimensions, $D_q(E)$. This requires analyzing the scaling of the $\text{IPR}_q$ with system size $N$, as described by Eq.~\eqref{eq:scaling_relation} for different values of $q$. To achieve this, we first compute $\text{IPR}_q(\Phi_k)$ values using Eq.~\eqref{eq:ipr_definition} for individual eigenstates $\Phi_k$ corresponding to the selected energies. These are obtained from SGs of varying generations, from $g = 5$ to $ g = 10$— corresponding to different system sizes $N$, as in SM Sec.~\ref{sec:sm_geometry}.

For a given generation $g$, in instances where multiple eigenstates $\Phi_k$ have energies $E_k$ satisfying the condition $|E_k - E| \le \Delta E$, their $\text{IPR}_q$ values are averaged. This procedure enhances statistical robustness, particularly for nearly degenerate energy levels:

\begin{equation}
    \label{eq:ipr_average} 
    \langle \text{IPR}_q(E) \rangle_{\Delta E} = \frac{1}{M_E} \sum_{k=1}^{M_E} \text{IPR}_q(\Phi_k),
\end{equation}

where $M_E$ is the number of such eigenstates $\Phi_k$ within the energy window $\Delta E$. 

From these averaged $\langle \text{IPR}_q(E) \rangle_{\Delta E}$ values, an effective, size-dependent mass exponent $\tau_q(E, N)$ is calculated for each system size $N$:
\begin{equation}
    \label{eq:tau_from_ipr}
    \tau_q(E, N) = -\frac{\ln\langle \text{IPR}_q(E) \rangle_{\Delta E} }{\ln N}.
\end{equation}

\begin{figure}[htb]
    \centering
    \subfigure[GS ($E\simeq -3.667$)]{%
        \includegraphics[width=0.3\textwidth]{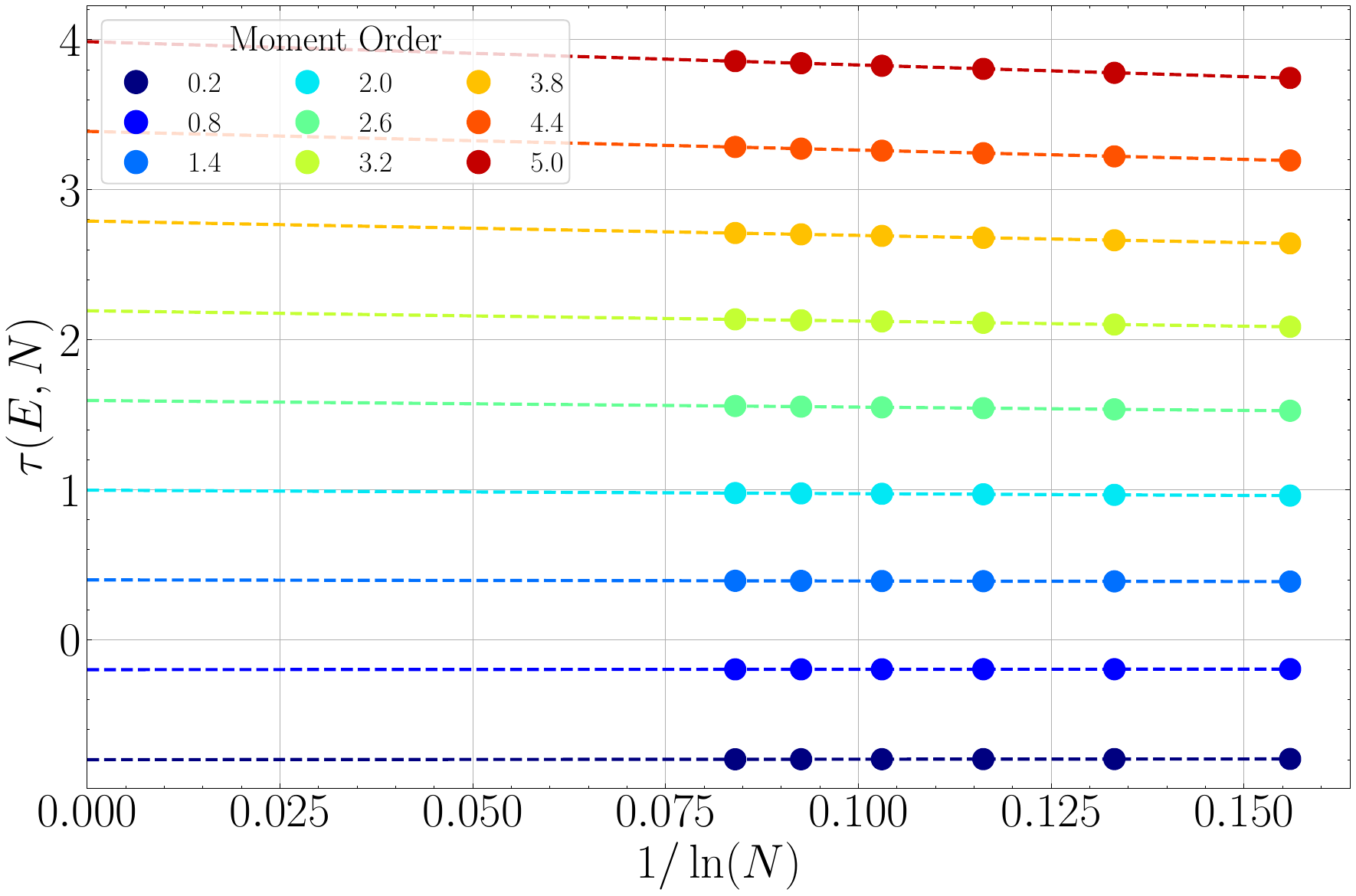}
        \label{fig:tqfinite_gs} 
    }
    \hspace{\fill} 
    \subfigure[NEM ($E\simeq 1.509$) ]{%
        \includegraphics[width=0.3\textwidth]{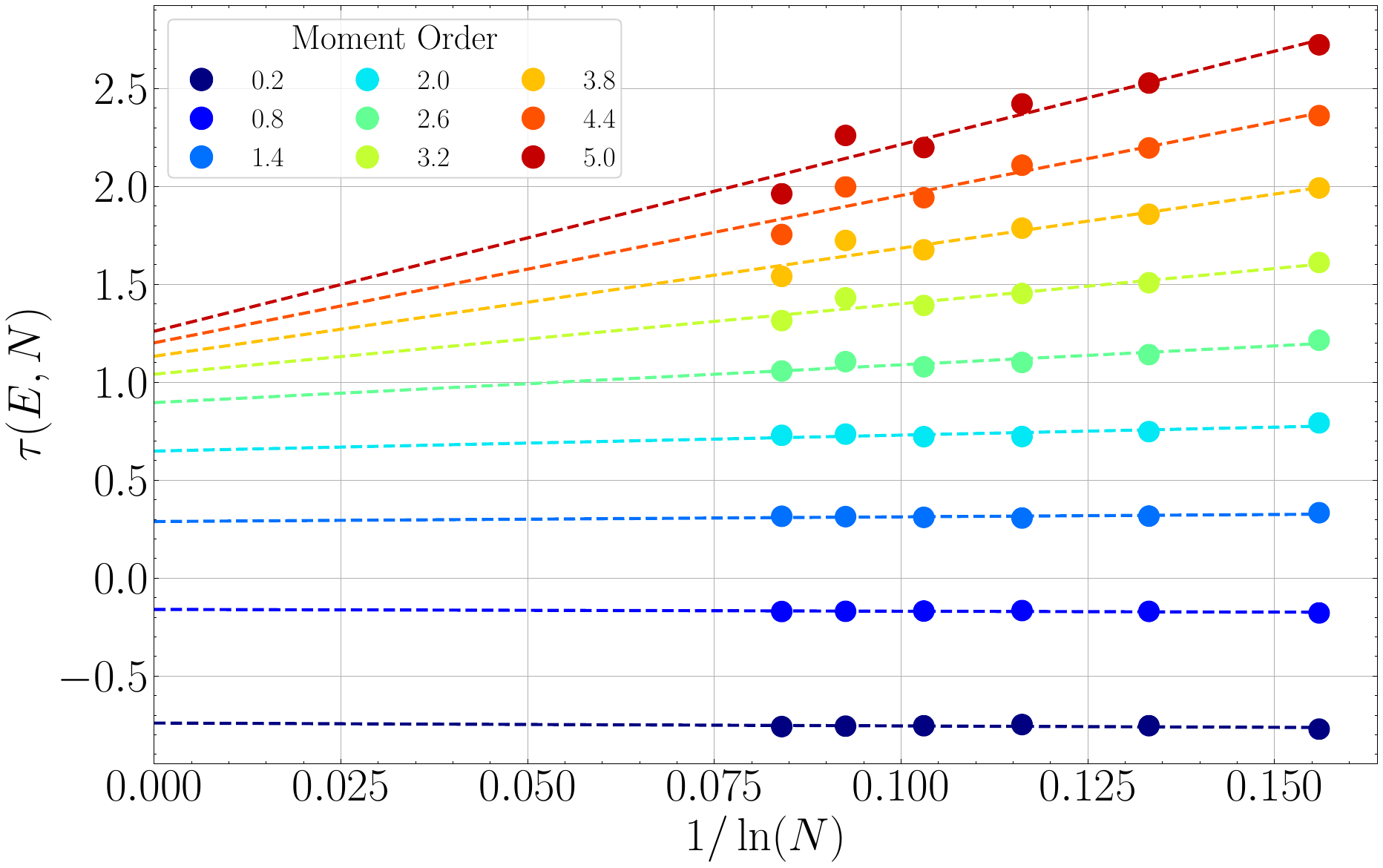}
        \label{fig:tqfinite_nem}
    }
    \hspace{\fill}
    \subfigure[LOC ($E\simeq -2.184$)]{%
        \includegraphics[width=0.3\textwidth]{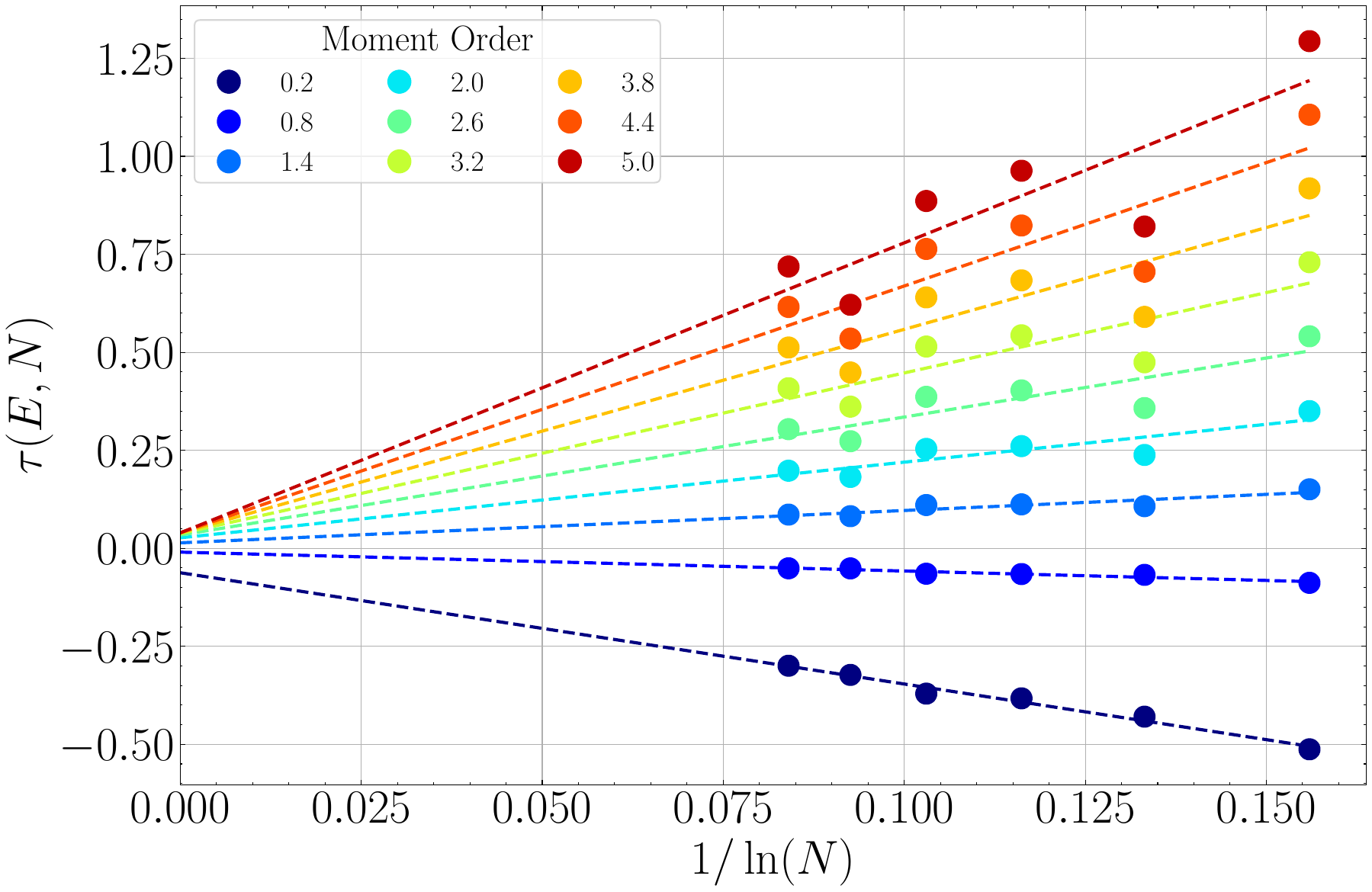}
        \label{fig:tqfinite_loc}
    }
    \caption{Finite-size scaling extrapolation of mass exponents for representative eigenstates.
    The panels show the effective mass exponents $\tau_q(E,N)$ plotted against $1/\ln N$ for various moment orders $q$. The thermodynamic limit value, $\tau_q(E)$, is obtained as the $y$-intercept ($1/\ln N \to 0$) of a linear fit (dashed lines) to the data points from different system sizes $N$ corresponding to Sierpiński gasket generations $g=5$ through $g=10$.\\
    (a) The ground state (GS), for which the extrapolated mass exponents $\tau_q(\text{GS})$ lead to $D_q(\text{GS}) = \tau_q(\text{GS})/(q-1)\simeq 1$, confirming the extended nature of the ground state. (b) The NEM state, considered as a candidate multifractal state. The exponents $\tau_q$ display a distinct non-linear dependence on $q$.
    (c) The candidate localized state (LOC): as the extrapolated $\tau_q \simeq 0$ for all $q > 0$.}
    \label{fig:tq_finite}

\end{figure}

To determine the exponent $\tau_q(E)$ in the thermodynamic limit, these finite-size values $\tau_q(E, N)$ are then extrapolated by plotting them against $1/\ln N$. The thermodynamic value $\tau_q(E)$—and consequently the fractal dimension $D_q(E) = \tau_q(E)/(q-1)$— is obtained as the intercept of a linear fit to these data in the limit $1/\ln N \to 0$.
The outcome is illustrated in Fig.~\ref{fig:tq_finite}.

\begin{figure}[h!]
    \includegraphics[width=0.7\linewidth]{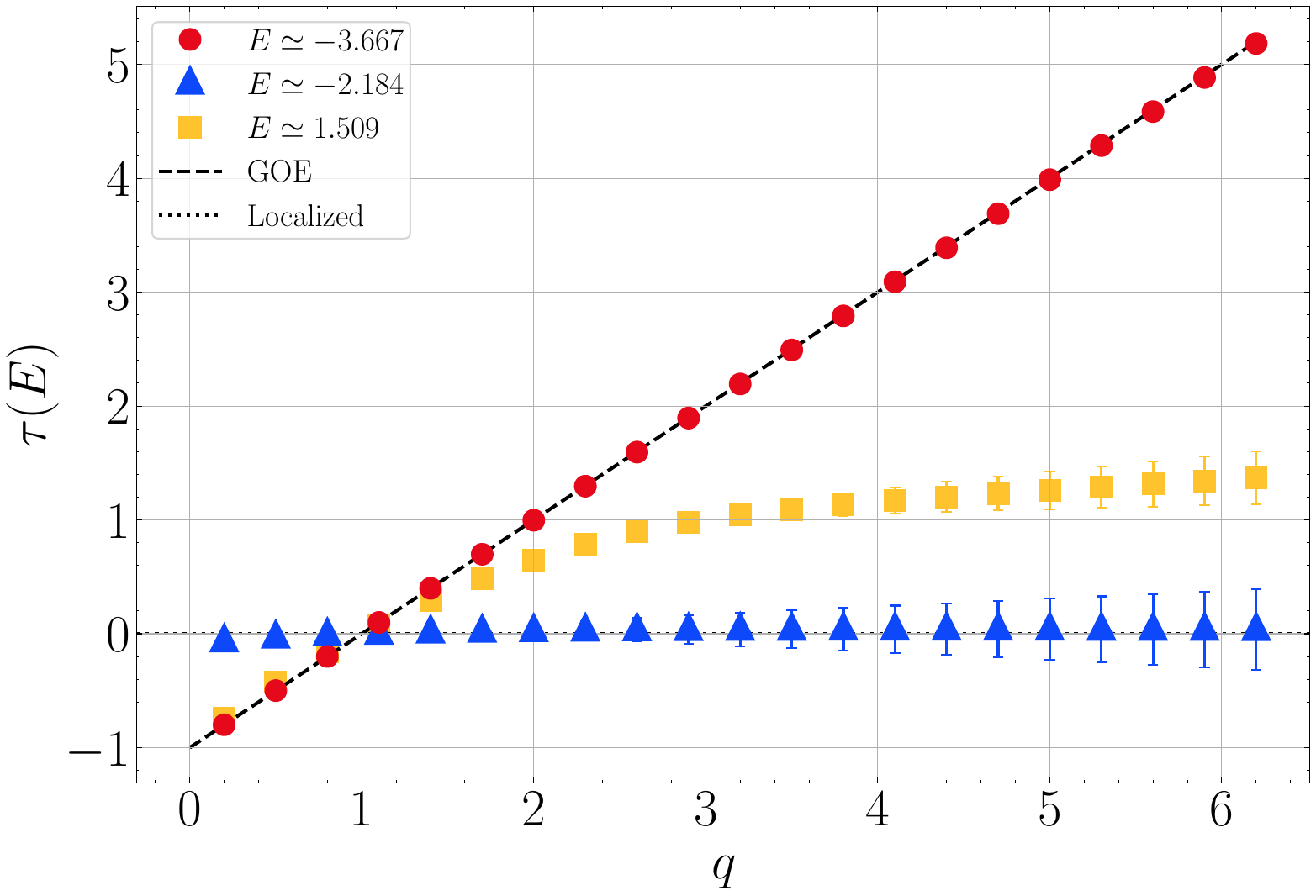}
    \caption{Extrapolated mass exponents $\tau_q(E)$ as a function of moment order $q$ for representative states on the Sierpiński gasket:
    Ground State (GS, $E\simeq -3.667$, red circles),
    Localized State (LOC, $E\simeq -2.184$, blue triangles), and
    Non-Ergodic Multifractal State (NEM, $E\simeq 1.509$, orange squares). Statistical errors are included; where not visible, the error bars are smaller than the marker size.
    The dashed black line indicates the behavior of ideal extended states ($\tau_q = q-1$, as for Gaussian Orthogonal Ensemble), while the dotted black line represents ideal localized states ($\tau_q = 0$).} 
    \label{fig:tq_therm}
\end{figure}

Finally, to provide a clear visual basis to rigorously classify the chosen states, in Fig.~\ref{fig:tq_therm}, we plot the mass exponents $\tau_q(E)$, obtained from the $y$-intercepts of these linear extrapolations, as a function of the moment order $q$. 

For the ground state (GS, $E\simeq -3.667$, red circles), the mass exponents $\tau_q(\text{GS})$ exhibit a clear linear dependence on $q$. This behavior closely follows the $\tau_q = q-1$ relation (dashed black line), characteristic of extended states, such as those in systems with Gaussian Orthogonal Ensemble (GOE) symmetry. The corresponding generalized fractal dimension $D_2(\text{GS})\simeq 0.997$, which supports its metallic-like nature.
In contrast, the state at $E\simeq -2.184$ (LOC, blue triangles) exhibits mass exponents $\tau_q(\text{LOC})$ that are approximately zero across the range of positive $q$ values. This $\tau_q\simeq 0$ behavior (dotted black line) is representative of exponential localization (see also SM, Sec.\ref{sec:sm_localization}), yielding generalized fractal dimensions $D_q(\text{LOC})\simeq 0$ for $q>0$. The specific value $D_2(\text{LOC})\simeq 0.017$ quantifies this strong confinement. The state at $E\simeq 1.509$ (NEM, orange squares) deviates significantly from both extended and localized limits. Its mass exponents $\tau_q(\text{NEM})$ display a distinct non-trivial dependence on $q$, a signature of multifractality; its fractal dimension $D_2(\text{NEM})\simeq 0.652$ further supports this. 
The collective analysis of these representative states, summarized by their distinct $\tau_q(E)$ signatures presented in Fig.~\ref{fig:tq_therm}, unequivocally demonstrates the coexistence of extended, localized, and multifractal eigenstates within the spectrum of the SG.

\begin{figure}[h!]
    \includegraphics[width=1\linewidth]{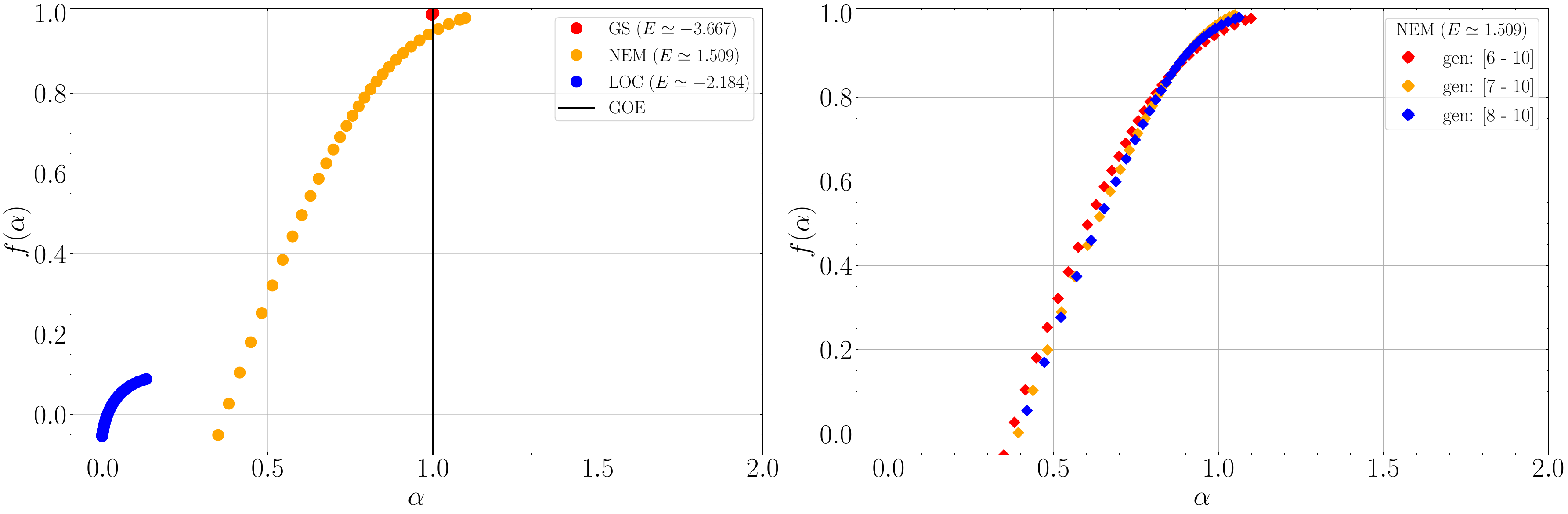}
    \caption{Singularity spectrum $f(\alpha)$ for representative states and its finite-size scaling. All spectra are calculated using the same range of moment orders $q$ as presented in Fig.~\ref{fig:tq_therm}. \\
    Left panel: Singularity spectrum $f(\alpha)$ for the representative states: extended (GS, red), localized (LOC, blue), and multifractal (NEM, orange). The spectrum for the Gaussian Orthogonal Ensemble (GOE, black line) is displayed as a reference for ideal extended states. Differences in the peak position and the width of the $f(\alpha)$ curve distinguish the localization classes.    
    Right panel: Finite-size scaling of $f(\alpha)$ for the NEM case study. Data comparing different generation ranges show agreement, supporting its multifractal nature.}
    \label{fig:gasket_singspec_canon}
\end{figure}

To further validate these findings, we examine the distribution of the singularity spectrum $f(\alpha)$ \cite{localization_mirlin}. We compute the spectra shown via the Legendre transform of the extrapolated mass exponent $\tau_q$ (see Methods), while an alternative analysis using the histogram method is presented in the SM, Sec.~\ref{sec:sm_singspec_supp}. The left panel of Fig.~\ref{fig:gasket_singspec_canon} compares the resulting hull of $f(\alpha)$ for the chosen representatives. The ground state (GS, red circles) exhibits a narrow $f(\alpha)$ curve sharply peaked near $\alpha_0 = 1$. This profile is in agreement with the theoretical expectation for fully extended states— as attested by the comparison with a realization of Gaussian Orthogonal Ensemble (GOE, thick black line). The insulating state (LOC, blue circles) satisfies the condition $f(\alpha \to 0) = 0$, a typical mark of localization in the singularity spectrum analysis. Finally, the representative non-ergodic multifractal state (NEM, orange circles) is distinguished by a broader, roughly parabolic $f(\alpha)$ curve. Its apex is visibly shifted to a value greater than one, and the finite width of the spectrum quantifies the strength of multifractality.

The robustness of the $f(\alpha)$ spectrum for the NEM state is also proved by its finite-size scaling, presented in the right panel of Fig.~\ref{fig:gasket_singspec_canon}. This anchored-window \cite{transport_luitz} analysis compares spectra from system size ranges sharing the largest available generation. The resulting spectra remain broad with substantial overlap, supporting the multifractal character of the state. On the other hand, the slow convergence, visible as a slight drift between the curves, is due to strong finite-size effects rather than a flow towards an ergodic state (see SM, Sec.~\ref{sec:sm_singspec_supp}).

Through the preceding finite-size scaling analysis of these exemplary cases, one can examine the localization properties across different groups of eigenstates, potentially unveiling a variety of degrees of multifractality. The SM Sec.~\ref{sec:sm_other_nems} also contains data for other NEMs, displaying different types of multifractal behavior, evident in their $f(\alpha)$ spectra—for example, in the width or the shift of the peak. Some of these NEMs retain metallic or insulating characteristics \cite{corr_cuevas, corr_mirlin} and may also exhibit spatial structures that resemble bulk and edge modes observed in artificial fractal systems \cite{fractals_kempkes, fractals_fischer}.

\subsection*{Two-point Correlator}
Building upon the characterization derived from eigenstate statistics, we now study the two-point density-density correlator \cite{corr_gorkov}, defined as:

\begin{equation}
    \label{eq:2eig_corr}
    \mathcal{F}(\omega, E) = \frac{ N \sum_r \sum_{i,j} |\psi_i(r)|^2 |\psi_j(r)|^2\, \delta\!\left(E - \lambda_i - \frac{\omega}{2} \right) \delta\!\left(E - \lambda_j + \frac{\omega}{2} \right) }{\sum_{i,j} \delta\!\left(E - \lambda_i - \frac{\omega}{2} \right) \delta\!\left(E - \lambda_j + \frac{\omega}{2} \right) } 
\end{equation}

It is worth emphasizing that, unlike Random Matrix Theory systems, which require ensemble averaging over disorder realizations, our system is deterministic and inherently disorder-free. Thus, $\mathcal{F}(\omega, E)$ directly measures the intrinsic correlations for our specific Hamiltonian between the densities of eigenstates $\psi_i$ and $\psi_j$ (with eigenvalues $\lambda_i$ and $\lambda_j$, respectively) that are separated by an energy $\omega$, with their average energy $(\lambda_i+\lambda_j)/2$ centered around $E$. 
This approach provides an experimentally accessible probe to reveal and discriminate between extended, localized, and NEM signature states, particularly through its characteristic scaling behavior at criticality.

\begin{figure}[h!]
    \includegraphics[width=0.7\linewidth]{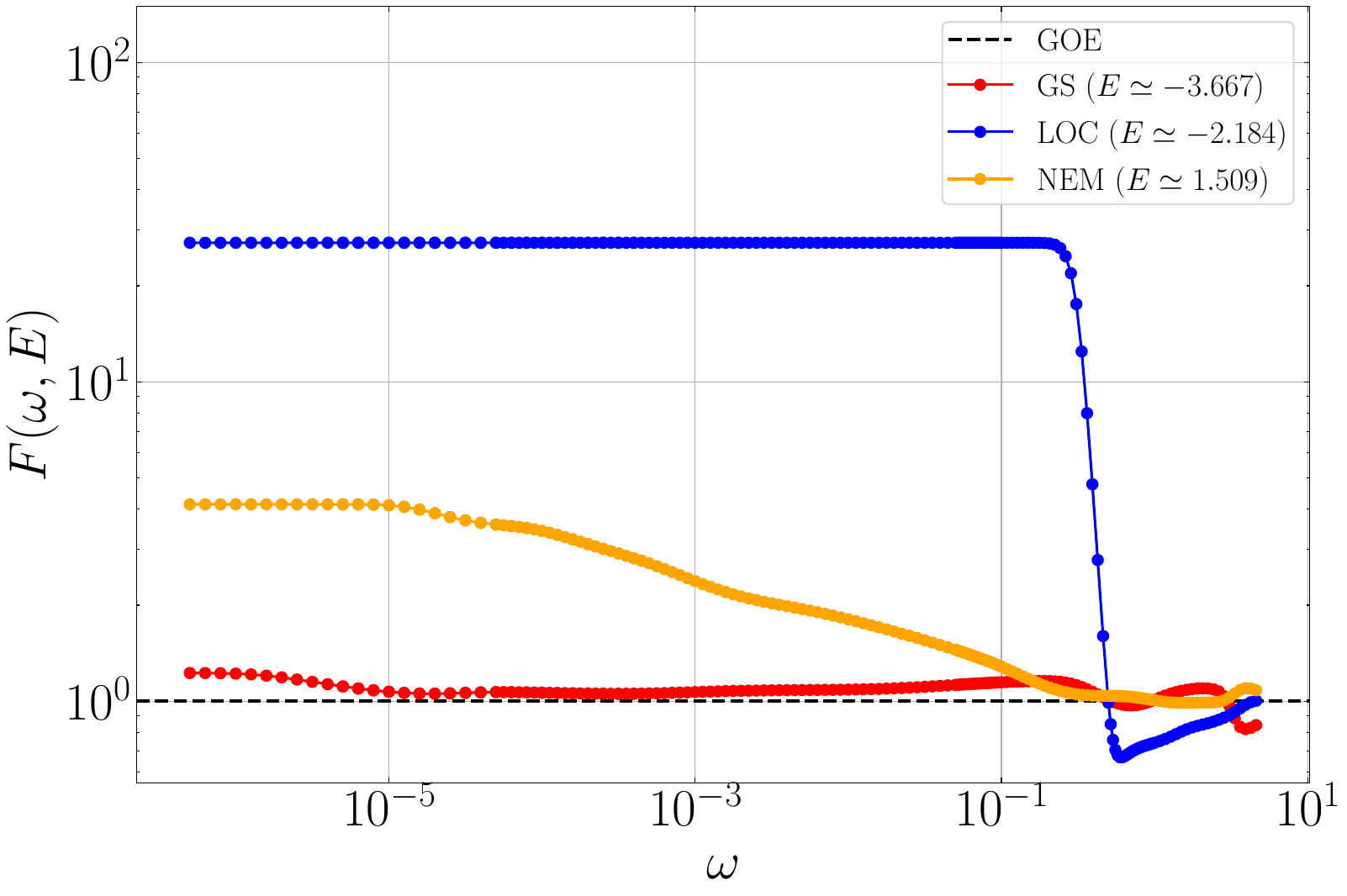}
    \caption{Log-log plot of the two-point density-density correlator $\mathcal{F}(\omega, E)$ versus energy separation $\omega$. Data are shown for three characteristic states from a $6$th generation Sierpiński gasket: the extended ground state (GS, $E \simeq -3.667$, red markers), a localized state (LOC, $E \simeq -2.184$, blue markers), and a non-ergodic multifractal state (NEM, $E \simeq 1.509$, orange markers). The dashed black line indicates the Gaussian Orthogonal Ensemble (GOE) reference for uncorrelated systems, corresponding to $\mathcal{F}(\omega,E) \approx 1$. While both extended and localized states exhibit broad plateaus at small $\omega$ indicating non-vanishing correlations, their respective heights are markedly different. Specifically, for the chosen localized state—situated in a spectral gap—the high plateau is a direct consequence of its triple degeneracy. In contrast, the NEM state features a narrower one (expected to become a singular point in the thermodynamic limit) followed by a characteristic power-law decay, $\mathcal{F}(\omega,E) \propto \omega^{-\mu}$, a fingerprint of multifractality. Lines are guides to the eye.}
    \label{fig:gasket_correlator_canon}
\end{figure}

Figure~\ref{fig:gasket_correlator_canon} illustrates $\mathcal{F}(\omega, E)$ for the previously selected representative extended, localized, and NEM states, computed for the $6$th generation SG. As predicted by theory, the two-point correlator for the extended state (red) and the localized state (blue) exhibits a plateau across several orders of magnitude at small energy separation in $\omega$. The plateau height reflects the degree of eigenstate overlap: relatively low and broad for extended states, due to their diffuse nature across the system, and significantly higher for localized states, which are concentrated in confined regions. At large energy separations in $\omega$, the correlator approaches the uncorrelated regime, with exponential decay for localized states. In this large-$\omega$ limit, $\mathcal{F}(\omega, E)$ tends towards unity (i.e., $\mathcal{F} \approx 1$), a value consistent with representative GOE statistics.

The two-point correlator for the NEM state (orange) displays a distinctly shorter plateau, followed by a characteristic power-law decay with energy separation $\omega$, described by $\mathcal{F}(\omega, E) \propto \omega^{-\mu}$, before approaching the uncorrelated regime. Power-law scaling over a broad $\omega$ range is a hallmark of multifractal states and reflects their inherent scale-invariance~\cite{corr_cuevas, corr_kravtsov}.

\begin{figure}[h!]
    \includegraphics[width=\linewidth]{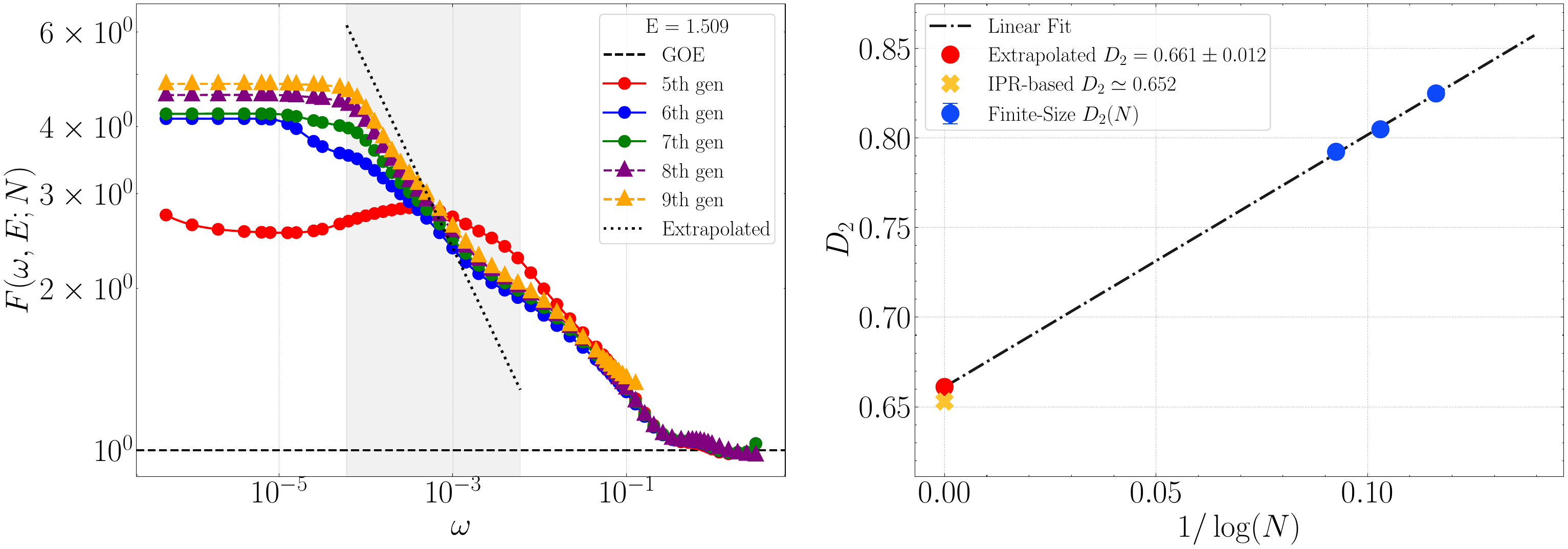}
    \caption{Finite-size scaling analysis of the two-point density-density correlator $\mathcal{F}(\omega, E; N)$ for the multifractal state at $E \simeq 1.509$. \\
    Left panel: Log-log plot of $\mathcal{F}(\omega, E; N)$ versus energy separation $\omega$ for Sierpiński gasket generations $g=5-7$ (circles, solid lines, exact diagonalization) and $g=8-9$ (triangles, dashed lines, subspace diagonalization).  
    The dotted black line shows the extrapolated power-law decay (with exponent~$\mu$ from the thermodynamic-limit value of $D_{2}$ in the right panel), while the gray shaded area marks the regime used for the fits. The dashed black line represents the GOE uncorrelated limit ($\mathcal{F}=1$).
    Right panel: Finite-size scaling extrapolation of the correlation dimension $D_2(N) \approx 1-\mu(N)$ against $1/\ln(N)$. Blue circles denote the $D_2(N)$ values derived from the decay exponents $\mu(N)$ for each generation. The dash-dot line is a linear fit, with its $y$-intercept (red circle) collapsing to thermodynamic limit value $D_2 = 0.661 \pm 0.012$. Statistical errors are included. The corresponding $D_2$ obtained from IPR analysis (orange cross) is displayed for comparison.}
    \label{fig:gasket_scaling_fit}
\end{figure}

To quantitatively verify this multifractal scaling and determine the correlation dimension $D_2$, we analyze the system-size dependence of $\mathcal{F}(\omega, E; N)$. The left panel of Figure~\ref{fig:gasket_scaling_fit} presents data for the NEM state at $E \simeq 1.509$ across multiple system sizes. As the system size $N$ increases, the $\mathcal{F}(\omega, E; N)$ curves exhibit improved convergence in their power-law region. 

Based on Chalker's scaling hypothesis for critical systems~\cite{corr_cuevas}, the exponent $\mu(N)$ is related to a size-dependent correlation dimension $D_2(N)$ by $\mu(N) \approx 1 - D_2(N)$. Thus, the value for $D_2$ in the thermodynamic limit is obtained by extrapolating these finite-size $D_2(N)$ values against $1/\ln N$. The finite-size scaling data are plotted in the right panel of Fig.~\ref{fig:gasket_scaling_fit} and yield the fractal dimension $D_2 = 0.661 \pm 0.012$. This value is in agreement, within the stated uncertainties, with the value $D_2 \simeq 0.652$ obtained independently from IPR-based eigenstate statistics. This consistency between two distinct methods strongly corroborates the emergence of multifractal states.

\subsection*{Robustness and Quantum Control}

So far, we have considered an idealized fractal structure. In reality, random defects existing on top of the regular fractal lattice are likely unavoidable. Such imperfections can disrupt the self-similar pattern, altering the nature of the NEM states observed in the pristine system. Therefore, the stability of these states against such disorder needs to be addressed.

For this purpose, we examine the characteristics of eigenstates in the SG lattice subjected to a varying number of additional random defects (ARDs). The Hamiltonian incorporating these random defects, $H_{\text{ARD}}$, is an extension of the original Hamiltonian $H$ from Eq.~\eqref{eq:hamiltonian}:

\begin{equation}
    \label{eq:hamiltonian_random}
    H_{\text{ARD}} = \!\! \sum\limits_{\substack{\langle i,j \rangle \\ i, j \notin (\mathcal{B} \cup \mathcal{R})}} \!\!t\, c^\dagger_i c_j \,\,+ \! \sum_{k \in (\mathcal{B} \cup \mathcal{R})}\!\! V\, c^\dagger_k c_k \,\, + \!\!\sum_{\substack{\langle i,k \rangle \\  k \in (\mathcal{B} \cup \mathcal{R})}} \!\! \overline{t}\, c^\dagger_i c_k + \text{h.c.}
\end{equation}

Here, the set $\mathcal{R}$ contains $N_\mathcal{R}$ sites chosen randomly from the active sites of the pristine SG (i.e., $\mathcal{R} \subset (\mathcal{L}\setminus\mathcal{B})$). Consistent with the structural defects defining the SG, these ARDs are modeled with the same on-site potential strength $V$ and reduced hopping $\overline{t}$.

According to the single-parameter scaling theory of Anderson localization~\cite{gang_four}, random geometric disorder leads to the localization of all electronic states for systems of dimensionality $d \le 2$. The SG, with its fractal dimension $d_f\simeq 1.585$, falls into this category, implying that its eigenstates would ultimately localize under such perturbations in the thermodynamic limit. Our aim here is more specific: to assess how the previously identified NEM states are affected by ARDs, and to what extent their characteristic properties survive in experimentally relevant finite-size systems. 

\begin{figure}[!h]
    \includegraphics[width=0.8\linewidth]{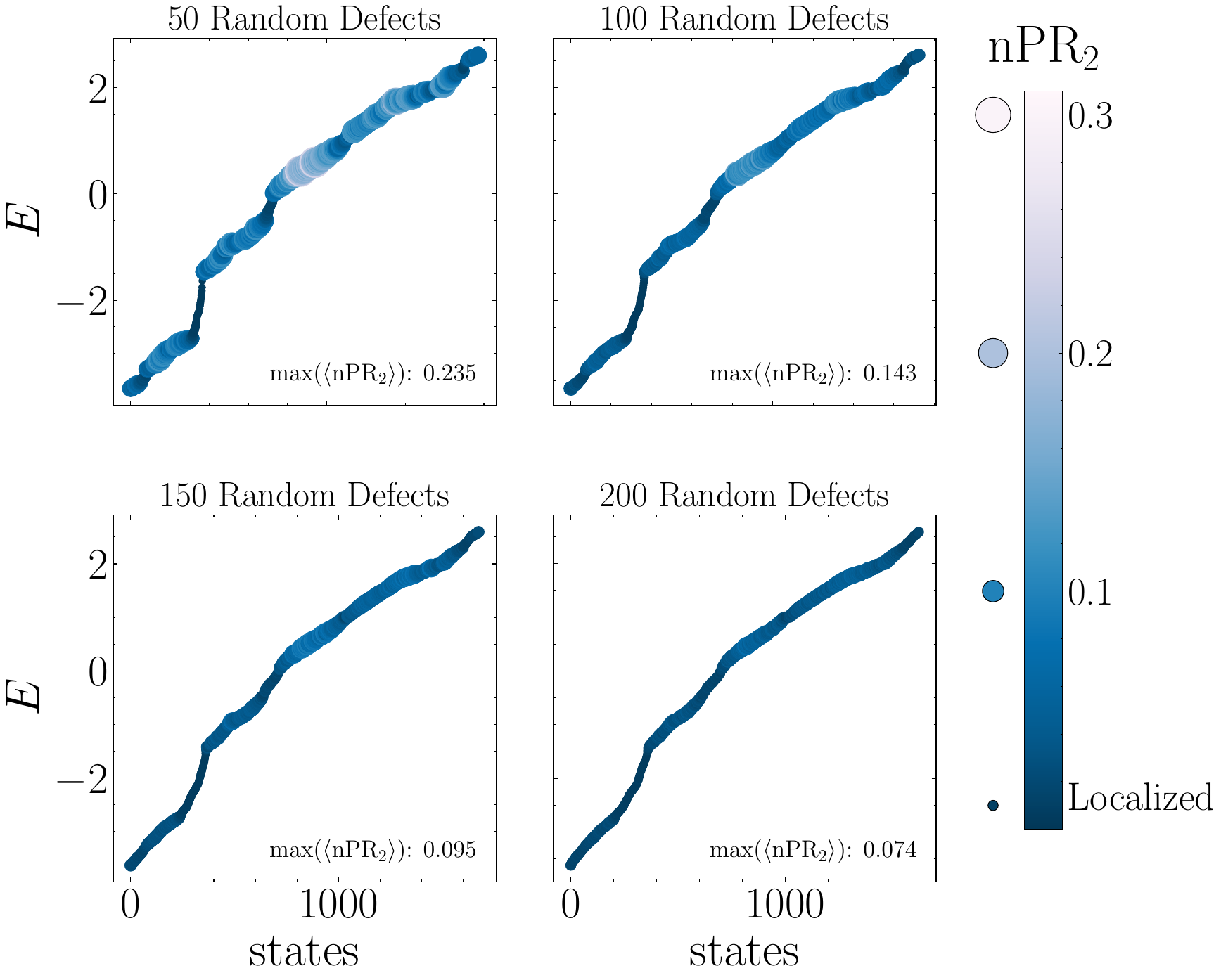}
    \caption{Energy spectrum of the single-particle tight-binding model on the 6th generation of Sierpiński gasket with additional random defects (ARDs), averaged over 10 disorder realizations. Configurations with 50, 100, 150 and 200 ARDs are analyzed. The color and marker size denote $\text{nPR}_2$ similarly to Fig.~\ref{fig:nPR_gasket}. In the lower right corner of each subplot, the maximum of $\text{nPR}_2$ is displayed, quantifying the tendency to localization as the number of defects increases.}
    \label{fig:gasket_nPR2_defects}
\end{figure}

Figure~\ref{fig:gasket_nPR2_defects} displays the $\text{nPR}_2$ distribution for all eigenstates of the $g=6$ SG as a function of $N_\mathcal{R}$, averaged over $p = 10$ geometric disorder realizations. As more ARDs are introduced, the average $\text{nPR}_2$ value across the spectrum, $\langle \text{nPR}_2 \rangle$, exhibits a clear decreasing trend, indicating a general shift towards stronger localization. Notably, critical states remain present across certain energy intervals, even when a substantial number of ARDs are present (e.g., $N_\mathcal{R}=200$, corresponding to approximately $10.7\%$ of the SG sites). Once again, the underlying hierarchical structure of the SG seems to delay a complete transition to universal Anderson localization, thereby preserving, in some states, the multifractal signature. 

The overall tendency of critical states to endure upon introducing random defects can be complemented by a more detailed analysis of the robustness of certain NEMs at specific energies. While disorder-induced spectral phenomena can facilitate the reorganization of NEMs through resonances and lead to new forms of multifractality, other NEMs maintain their spatial distribution without significant alteration. 

\begin{figure}[!h]
    \includegraphics[width=1\linewidth]{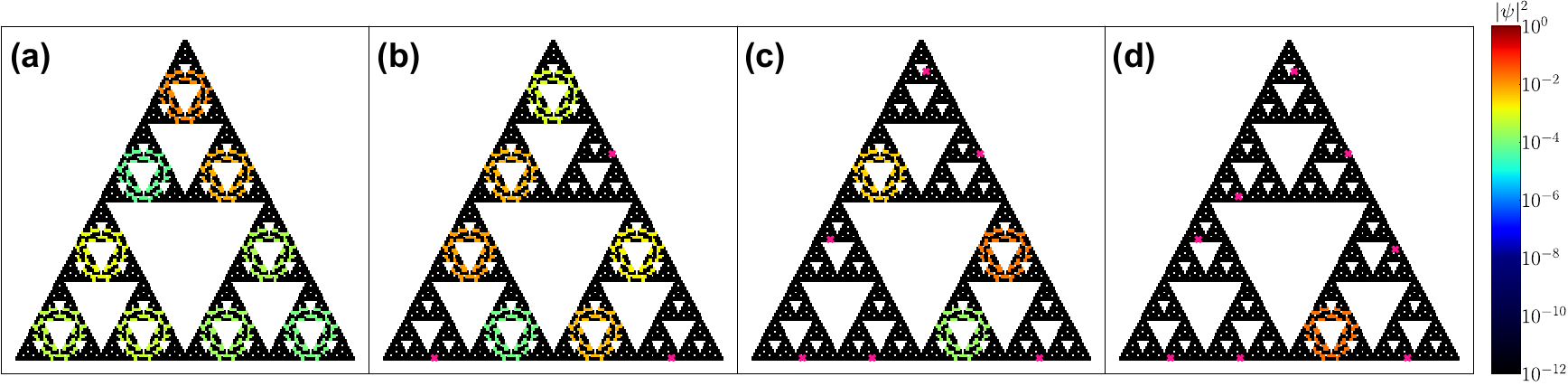}
    \caption{Amplitude profile of the state at energy level $E = 1.000$. Profiles (a) in the original model without defects, and (b-d) with additional defects (marked by pink crosses) placed at three, six, and eight targeted sites, respectively. Each defect, positioned within its respective cluster, completely blocks information propagation in that region, while causing lossless redistribution of information among the remaining clusters. The colorbar indicates $|\psi|^2$, with red corresponding to high amplitude values and black to nearly zero ones.}
    \label{fig:mcs_single_defects}
\end{figure}

A remarkable example is the eigenstate at $E\simeq1.000$ depicted in Fig.~\ref{fig:mcs_single_defects}a, for the $6$th generation SG. Characterized as multifractal (details in SM Sec.~\ref{sec:charact_multif}), its unique wavefunction intensity profile is sparse, confined to several disconnected groups of sites. These clusters are predominantly located near the internal edges of the SG structure, between the third and fourth hierarchical levels. More importantly, this distinct spatial profile is reminiscent of compact localized states \cite{cls_schmelcher} and of topologically equivalent nodes from the graph theory \cite{node_kochergin}, suggesting these types of states can hold considerable potential for applications in quantum information storage and transfer~\cite{cls_rontgen, cls_lazarides}.

These isolated wavefunction clusters, indeed, remain largely unaffected by ARDs unless the defects are placed directly onto their supporting lattice sites.
When defects are systematically introduced on sites within these specific clusters, the degeneracy is lifted, and the wavefunction amplitude is redistributed among the remaining, unaffected sets, with minimal leakage to other non-resonant sites (Fig.~\ref{fig:mcs_single_defects}b-d). This targeted response demonstrates a pathway for controlled manipulation of the wavefunction's intensity, offering avenues for precise quantum control and motivating exploration of related phenomena, such as quantum scars. Moreover, as elaborated in the SM Sec.~\ref{sec:sm_robustness}, even when a larger number of ARDs significantly perturbs the original profile, these disturbances can trigger new quantum interference effects. Consequently, amplitude from a disrupted set can re-cohere and stabilize around lower hierarchical levels of the fractal, giving rise to novel multifractal clustered patterns. The prospect of experimentally realizing these structurally unique and potentially applicable states—for instance, in atomic quantum simulators~\cite{stm_jolie, stm_sierda} through precise chemical potential tuning—makes future inquiry particularly compelling.

\section*{Discussion}

With this work, we continue the study of quantum phenomena in artificial fractal geometries, focusing on the Sierpiński gasket. We have shown that in clean fractals, without any conventional randomness, non-ergodic multifractal states naturally emerge, complementing prior studies on wavefunction localization \cite{fractals_kempkes}, quantum transport \cite{transport_xu}, eigenstate statistics \cite{fractals_yao}, and energy spectra \cite{fractals_levels_iliasov} in fractals. We performed a multi-tool analysis of the properties of the SG spectrum and the features of individual NEMs. In addition, we provided sensitive indicators that distinguish their diverse attributes, confirming that these states are robust and generic rather than rare (SM, Sec.~\ref{sec:sm_robustness}).

Among other findings, we demonstrated a particular class of NEM states (e.g., at $E \simeq 1.000$) which, to our understanding, is unique to fractal geometries. These states respond in a highly symmetric and structured way to the introduction of geometric disorder, making this responsiveness an interesting platform for quantum control. We speculate that the displayed properties of such NEMs are analogous to those of compact localized states  \cite{cls_schmelcher, cls_rontgen} and that dispersionless wave propagation may be accomplished. If confirmed, this could enable controlled information flow between targeted regions, offering new approaches to quantum control. 

Not to limit ourselves to the presented model, we generalized our findings beyond the basic SG model with only nearest-neighbor hopping. In doing so, we included additional interactions, such as next-nearest-neighbor (NNN) hopping terms and revealed that NEMs still emerge under these conditions and exhibit robustness even as the NNN hopping strength is gradually increased (SM, Sec.~\ref{sec:sm_nnn_hopping}). 

Our exploration also covered other fractal structures, notably the negative Vicsek-saltire fractal (SM, Sec.~\ref{sec:sm_vicsek}). In this geometry, vacancies, rather than the lattice sites themselves, form a fractal pattern with a relatively low Hausdorff dimension ($d_f \simeq 1.46$).
This particular example was chosen because its creation demands fewer defects than the SG or the Sierpiński carpet, making its experimental realization easier. While the large translationally invariant "clean" sub-domains support extended states, critical states still emerge in certain parts of the spectrum.

We suggest that artificial fractal structures—systems that have already been created in a lab—are a natural platform to create and study multifractal states experimentally, which, to our knowledge, has so far been possible only in doped semiconductors \cite{stm_richardella} and still remains an ongoing challenge. Our numerical results, particularly on the two-point density-density correlation function $\mathcal{F}(\omega, E)$, provide a direct method to distinguish NEMs using scanning probe microscopy techniques (recent research has also demonstrated applicability of similar methods in acoustic metamaterials \cite{acoustic_zheng} and photonic systems \cite{fractals_biesenthal, photonic_li}).

The signatures of multifractal states that we have analyzed here lay the natural ground for extending this research.
In particular, moving beyond the single-particle framework, future studies could reveal connections to timely many-body phenomena, such as many-body localization \cite{mbl_manna}, Hilbert space fragmentation \cite{mbl_pietracaprina}, and quantum scars \cite{scars_serbyn, scars_wang}.

\section*{Methods}
\subsection*{Singularity Spectrum}

The multifractal nature of an eigenstate $\Phi$ can be further characterized by its singularity spectrum $f(\alpha)$. This function describes the fractal dimension of the set of sites $i$ where the local wavefunction intensity $|\langle i | \Phi \rangle|^2$ scales as $N^{-\alpha}$ in the large system size limit ($N \to \infty$)~\cite{singularity_jensen, corr_kravtsov, multifractality_mirlin}. In principle, $f(\alpha)$ can be determined from site-specific scaling exponents:
\begin{equation}
    \alpha_i = -\frac{\ln(|\langle i | \Phi \rangle|^2)}{\ln N},
    \label{eq:singspec_histogram}
\end{equation}
but for finite systems this direct computation may not converge to a size-independent form. We therefore employ an indirect method, which relies on the generalized inverse participation ratios, $\text{IPR}_q(\Phi)$ as in Eq.~\eqref{eq:ipr_definition}, and their established scaling behavior, $\text{IPR}_q(\Phi) \sim N^{-\tau_q(\Phi)}$, characterized by the mass exponents $\tau_q$.

The energy-resolved mass exponents $\tau_q(E)$, obtained via the extrapolation procedure detailed in the main text, are used to compute the singularity spectrum $f(\alpha,E)$ through a Legendre transform~\cite{sizescaling_rodriguez}:

\begin{align}
  \alpha_q(E) &= \frac{d \tau_q(E)}{dq}, \label{eq:alpha_q_definition_methods} \\
  f(\alpha_q(E), E) &= q \alpha_q(E) - \tau_q(E). \label{eq:f_alpha_definition_methods}
\end{align}

The parameter $q$ acts as a ``magnifying glass,'' emphasizing different parts of the measure $|\langle i | \Phi \rangle|^2$, analogous to an inverse temperature in statistical mechanics~\cite{multifractality_fyodorov}.

This procedure generates the convex hull of the singularity spectrum, whose characteristic shape allows for the classification of states. Extended states are distinguished by an $f(\alpha)$ spectrum that, in the ideal limit, is a single point at $(\alpha=1, f(\alpha)=1)$. Numerically, this is observed as a narrow peak around this point, where the peak value is the capacity dimension $f(1) = D_0 = 1$. 
Localized states, with wavefunctions confined to $\mathcal{O}(1)$ sites, exhibit generalized fractal dimensions $D_q \approx 0$ for $q>0$, implying $\tau_q \approx 0$ for $q>0$. Consequently, their $f(\alpha)$ spectrum effectively collapses to the point $\alpha_0 = 0$ with $f(\alpha_0) = 0$. For these states, the capacity dimension $D_0 = -\tau_0 \approx 0$.
Finally, multifractal states are characterized by a broader, typically convex $f(\alpha)$, independent of the size of the system and spanning a finite range of $\alpha$ values. The maximum of this curve occurs at $\alpha_0 = \alpha_{q=0}$ and it can deviate from 1. The peak's value is the dimension $f(\alpha_0)=D_0$, with $0 < D_0 < 1$; this is conceptually equivalent to the fractal dimension of the support set as obtained by a box-counting methodology. The overall width of the $f(\alpha)$ spectrum quantifies the degree of multifractality. The universality of certain symmetry relations for $f(\alpha)$ \cite{multifractality_mirlin}, such as $f(1+\alpha) = f(1-\alpha)+\alpha$, is restricted to specific systems or observables~\cite{nonergodic_tang, sing_bilen} and is not generally assumed for wavefunctions in our context. 

\subsection*{Two-eigenstate Correlator}
Following Berezinskii and Gorkov \cite{corr_gorkov}, the two-point density correlator, at energy $E$ for an arbitrary pair of points $(r,r')$, is generally defined as:

\begin{equation}
    \label{density_two}
     \langle \rho_E(r)\rho_{E+\omega}(r') \rangle = \left\langle \dfrac{1}{n(E)} \sum_{i,j} |\psi_i(r)|^2 |\psi_j(r')|^2 \delta(E - \lambda_i)\delta(E - \lambda_j +\omega) \right\rangle,
\end{equation}

where angle brackets $\langle \dots \rangle$ denote an average over disorder realizations, $n(E)$ is the density of states, and $\omega$ is the energy difference between the two states.

Within this formalism, for extended states, the wavefunctions are considered spread over all $N$ sites of the system, and averaged quantities may scale accordingly. In the localized case, wavefunctions have their support in a finite volume proportional to the square of the characteristic localization length, and the density correlator is generally dominated— apart from a nonsingular contribution—by a pole singularity: 

\begin{equation}
    \label{density_loc}
     \langle \rho_E(r)\rho_{E+\omega}(r') \rangle \sim \langle A_E(r - r') \rangle \delta(\omega),
\end{equation}

where the spectral function is:

\begin{equation}
    A_E(r - r') = \dfrac{1}{n(E)} \sum_{i}|\psi_i(r)|^2 |\psi_i(r')|^2 \delta(E - \lambda_i).
\end{equation}

The sum of this spectral function over coincident points yields the average Inverse Participation Ratio, $\langle \text{IPR}_2 \rangle_E$. Thus, this quantity is directly connected to the correlation dimension $D_2(E)$ via the characteristic scaling \newline $\langle \text{IPR}_2 \rangle_E \sim N^{-D_2(E)}$ (see Eqs.~\eqref{eq:ipr_definition}-\eqref{eq:scaling_relation}).

Such two-point density correlations have been widely applied in the study of disordered quantum systems \cite{corr_chalker88, corr_chalker90, corr_kravtsov}, including those lacking translational invariance \cite{multifractality_cugliandolo, fibonacci_andreanov}, and also to investigate intriguing phenomena such as mutual avoidance of eigenstates \cite{corr_cuevas}. Since the system considered in this work—a quantum system on a geometric fractal—is deterministic and possesses no quenched disorder, we directly compute a spectral correlation function $\mathcal{F}(\omega, E)$ without ensemble averaging. The form relevant to our analysis is:

\begin{equation}
    \label{eq:2eig_corr_methods}
    \mathcal{F}(\omega, E) = \frac{N \sum_r \sum_{i,j} |\psi_i(r)|^2 |\psi_j(r)|^2\, \delta\!\left(E - 
 \lambda_i - \dfrac{\omega}{2} \right) \delta\!\left(E - 
 \lambda_j + \dfrac{\omega}{2} \right)}{ \sum_{i,j} \delta\!\left(E - 
 \lambda_i - \dfrac{\omega}{2} \right) \delta\!\left(E -  \lambda_j + \dfrac{\omega}{2} \right) }.
\end{equation}

This function effectively quantifies overlaps between eigenstates separated by energy $\omega$.
For numerical simulations, each $\delta$-function is smeared by a normalized Gaussian, and $\mathcal{F}(\omega,E)$ is evaluated on a predefined logarithmically spaced $\omega$ grid. The Gaussian factors are applied separately to the numerator and the denominator before taking their ratio. To improve statistics, as in the IPR analysis, we average over a narrow energy window centered at $E$ by including all eigenstate pairs whose mean energy $(\lambda_i+\lambda_j)/2$ lies inside that window and averaging their contributions. This local energy averaging is performed within a single deterministic spectrum and does not involve any averaging over disorder.

For ideal extended states, such as those of the Gaussian Orthogonal Ensemble, $\mathcal{F}(\omega, E) = 1$ across the spectral range, indicating effectively uncorrelated eigenfunctions. In more realistic extended systems, like typical metals, a broad plateau where $\mathcal{F}(\omega, E) > 1$ often appears. The plateau is often bounded by a cutoff energy~\cite{repulsion_altshuler}—the Thouless energy $E_{Th}$—at the scale of the lattice spacing, beyond which the system returns to the uncorrelated regime. 
As the system transitions towards localization, these correlations become more pronounced for small energy separations, causing the correlation plateau to grow in height while narrowing. This high, narrow feature is then followed by the signature exponential decay of correlations that recovers the uncorrelated regime at larger values of $\omega$.

For a multifractal state, the correlation plateau is present only within a narrow energy interval, and the Thouless energy is expected to shrink to zero in the thermodynamic limit. In this case, the uncorrelated limit at large values of $\omega$ is preceded by a power-law decay $\mathcal{F}(\omega, E) \sim \omega^{-\mu}$, where the exponent $\mu$ is related to the correlation dimension $D_2$ by  $\mu \approx 1-D_2(E)$ \cite{corr_chalker88, corr_cuevas}. The system size $N$ influences $\mathcal{F}(\omega, E)$ for multifractal states, providing avenues to determine fractal dimensions. For instance, the plateau height at small $\omega$ is expected to scale as $\mathcal{F}(0, E) \propto N^{1-D_2(E)}$. More generally, scaling theory predicts that the $\mathcal{F}(\omega,E)$ curves for different $N$ should collapse under rescaling \cite{corr_cuevas}, however we do not apply this in the present work. Instead, we extract the decay exponents $\mu(N)$ directly from the raw $\mathcal{F}(\omega,E;N)$ data, and determine $D_{2}$ via the relation $\mu(N) \approx 1 - D_{2}(N)$.

Finite-size corrections, not considered here, could be included to improve the extrapolation of $D_{2}(E)$ to the thermodynamic limit.

\section*{Availability of data}
Data are available from the corresponding author upon reasonable request.

\section*{Acknowledgements}
We are grateful to C. Morais Smith, M.R. Slot, A. Moustaj, M.Conte and Y. in 't Veld for insightful discussions. We would also like to thank I.M. Khaymovich for valuable correspondence. This work was supported by the European Research Council (ERC) under the European Union's Horizon 2020 research and innovation program, grant agreement 854843-FASTCORR. 

%% file: arxiv_supplementary.tex
\title{Supplementary material}
\maketitle

\section{Sierpiński Gasket Geometry}
\label{sec:sm_geometry}

The site counts of the Sierpiński gasket (SG), for different generations ($g$) used in this study, are detailed below.

The total number of sites $N_\mathcal{L}(g)$ of the underlying triangular lattice region is calculated iteratively. This calculation utilizes an auxiliary term, $C(g)$, which itself follows the recurrence:
\begin{equation*}
C(g) = 2 C(g-1) - 3.
\end{equation*}

The recurrence for $N_\mathcal{L}(g)$ is then:
\begin{equation*}
 N_\mathcal{L}(g) = 4 N_\mathcal{L}(g-1) - C(g),
\end{equation*}
with an initial value of $N_\mathcal{L}(1) = 10$ for the elementary triangular block.

Similarly, the number of active sites $N_g(g)$ forming the gasket structure itself grows with generation $g$ according to:
\begin{equation*}
N_g(g) = 3 N_g(g-1) - 3,
\end{equation*}
with $N_g(1)=9$ active sites in the $g=1$ building block (as illustrated in the main text, MT Fig.~\ref{fig:triangle_to_gasket}).

Finally, the number of defect sites $N_\mathcal{B}(g)$ (i.e., voids introduced by the SG construction) is the difference:
\begin{equation*}
N_\mathcal{B}(g) =  N_\mathcal{L}(g) - N_g(g).
\end{equation*}
Table~\ref{tab:sg_geometry_supp} lists these site counts for the generations relevant to this work.

\begin{table}[htbp]
\centering
\label{tab:sg_geometry_supp}
\begin{tabular}{crrr}
    \toprule
    Generation ($g$) & \quad Total Sites ($N_\mathcal{L}$) & \quad Defect Sites ($N_\mathcal{B}$) & \quad Gasket Sites ($N_g$)\\
    \midrule
    4 & 325    & 121    & 204    \\
    5 & 1225   & 616    & 609    \\
    6 & 4753   & 2929   & 1824   \\
    7 & 18721  & 13252  & 5469   \\
    8 & 74305  & 57901  & 16404  \\
    9 & 296065 & 246856 & 49209  \\
    10& 1181953& 1034329& 147624 \\
    \bottomrule
\end{tabular}
\caption{$N_\mathcal{L}$ represents the total number of sites in the underlying triangular lattice region considered for a given generation $g$, $N_\mathcal{B}$ is the number of defect sites introduced (voids), and $N_g = N_\mathcal{L} - N_\mathcal{B}$ is the number of sites in the resulting gasket structure.}
\end{table}

\section{On the Hierarchy of Spectral Gaps}
\label{sec:sm_hierarchy}
The energy spectrum of the SG, as displayed in Fig.~\ref{fig:nPR_gasket} of the MT, is generated via an iterative map. For the simplified case of a tight-binding Hamiltonian with only nearest-neighbor hopping $t$: 

\begin{equation*}
H = -t \sum_{\langle i,j \rangle} c_i^\dagger c_j, 
\end{equation*}

the (shifted and scaled) energy levels $x_g$ of generation $g$ are related to those of generation $g-1$ by the recurrence relation \cite{fractals_kadanoff}:

\begin{equation}
\label{eq:spectrum_map}
x_{g} = \pm\sqrt{\gamma - x_{g-1}}
\end{equation}

with $\gamma = 15/4$ for the standard SG case. This iterative process naturally defines a hierarchy: an initial energy interval containing the spectrum is recursively divided into subintervals, leading to the formation of a self-similar structure of bands and gaps. The system described by Eq.~\eqref{eq:spectrum_map} can be approximated by a piece-wise linear model, which allows the retrieval of the full distribution of all gap lengths, as reported in \cite{fractals_levels_iliasov}.

\section{Additional Results for the Representative Extended, Localized, and Non-Ergodic Multifractal States}

\subsection{Amplitude Profiles}
\begin{figure*}[!htb]
    \includegraphics[width=\textwidth]{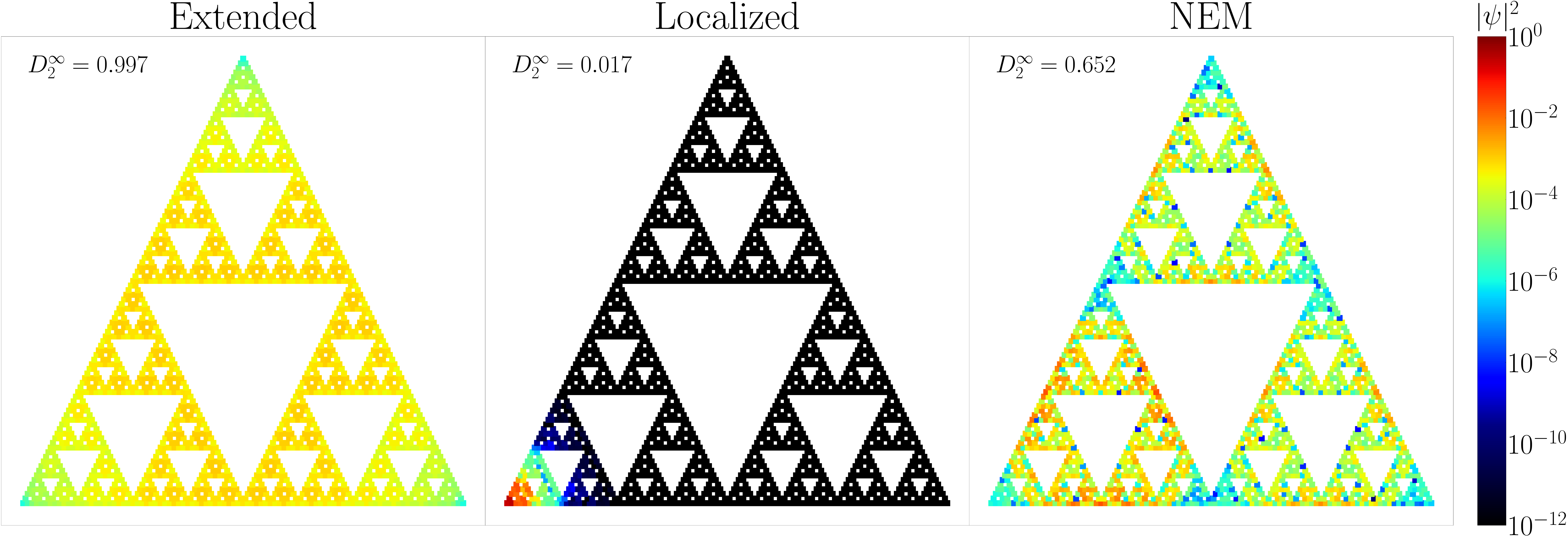}
    \caption{Intensity profiles of the representative ergodic ($E \simeq -3.667$), localized ($E \simeq -2.184$), and NEM ($E \simeq 1.509$) states. Red represents high intensity, black negligible intensity. For the sake of simplicity, the defects that reshape the triangular lattice into a SG are shown as voids. In the ergodic phase, the intensity is homogeneously distributed across all sites, indicating delocalization. The localized state exhibits a strong concentration of intensity in a small region, in this case near the corner. 
    The NEM state has a heterogeneous intensity profile with strong fluctuations that differs from either extended or localized states, and repeats across all scales.}
    \label{fig:sm_gasket_amplitude_canon}
\end{figure*}

The spatial distributions of the representative extended, localized, and non-ergodic multifractal (NEM) states, as analyzed in the MT, are visualized through their wavefunction intensity profiles, in Fig.~\ref{fig:sm_gasket_amplitude_canon}. 

\subsection{Spatial Decay Profile of the Localized State}
\label{sec:sm_localization}
\begin{figure*}[!htb]
    \includegraphics[width=0.7\textwidth]{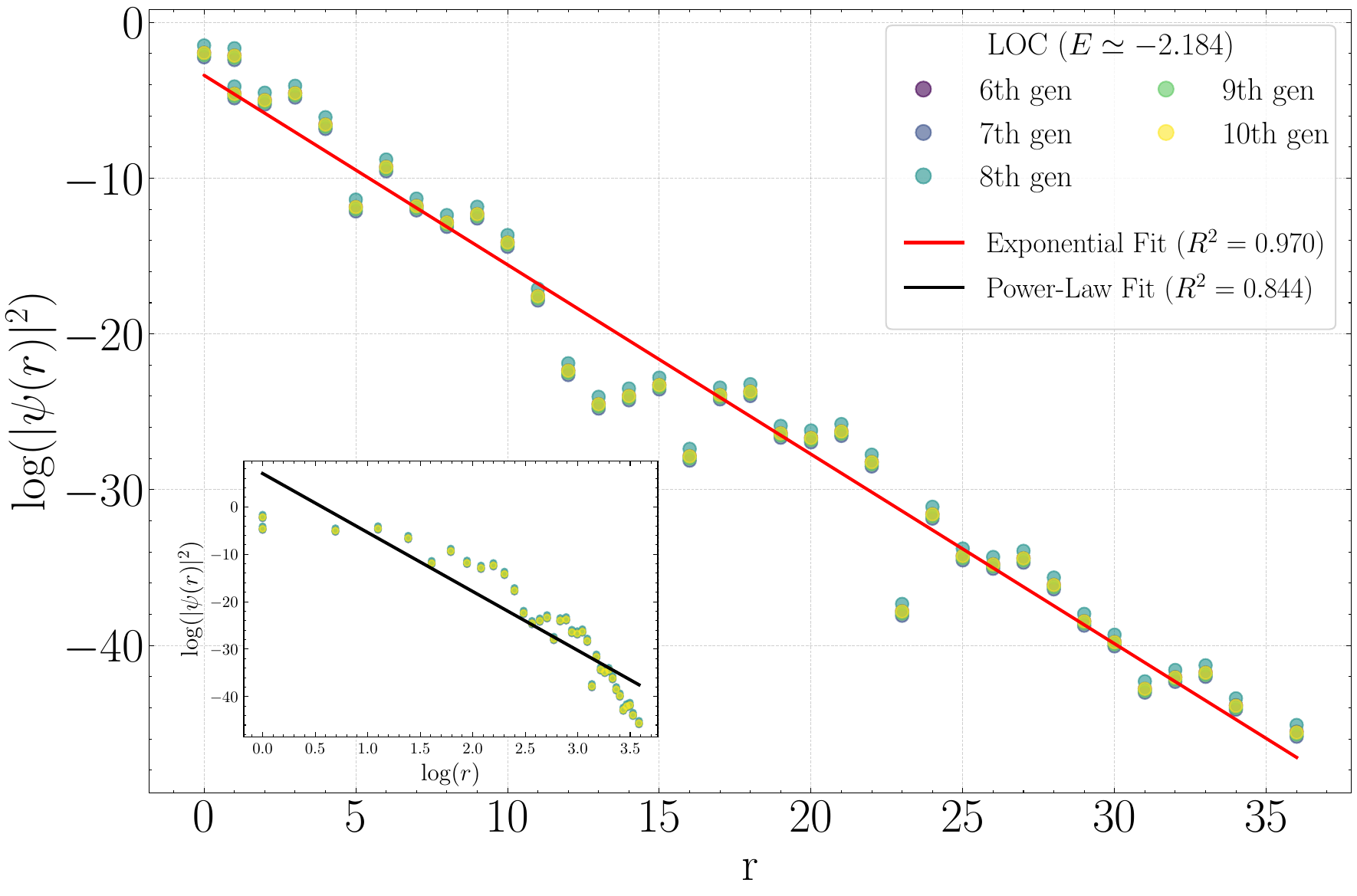}
    \caption{Decay profile analysis for the representative localized state at $E \simeq -2.184$. The main panel (semi-log scale) and inset (log-log scale) test for exponential and power-law decay, respectively. The resulting coefficients of determination are $R^2 = 0.970$ for the exponential fit and $R^2 = 0.844$ for the power-law fit, favoring the former.}
    \label{fig:localization_fit}
\end{figure*}

To quantitatively characterize the spatial confinement of the representative localized state ($E \simeq -2.184$), we fit its wavefunction intensity profile, $|\psi(r)|^2$, against exponential and power-law decay models as a function of distance $r$ from its peak. To distinguish between the models, we perform a linear regression on the data using both semi-log and log-log scales. A comparison of the resulting coefficients of determination ($R^2$) shows a clear preference for the exponential model. For statistical robustness, the fit presented in Fig.~\ref{fig:localization_fit} aggregates data from generations 6 to 10.

\subsection{Supplemental Analyses of the Singularity Spectrum}
\label{sec:sm_singspec_supp}
\begin{figure*}[!htb]
    \includegraphics[width=\textwidth]{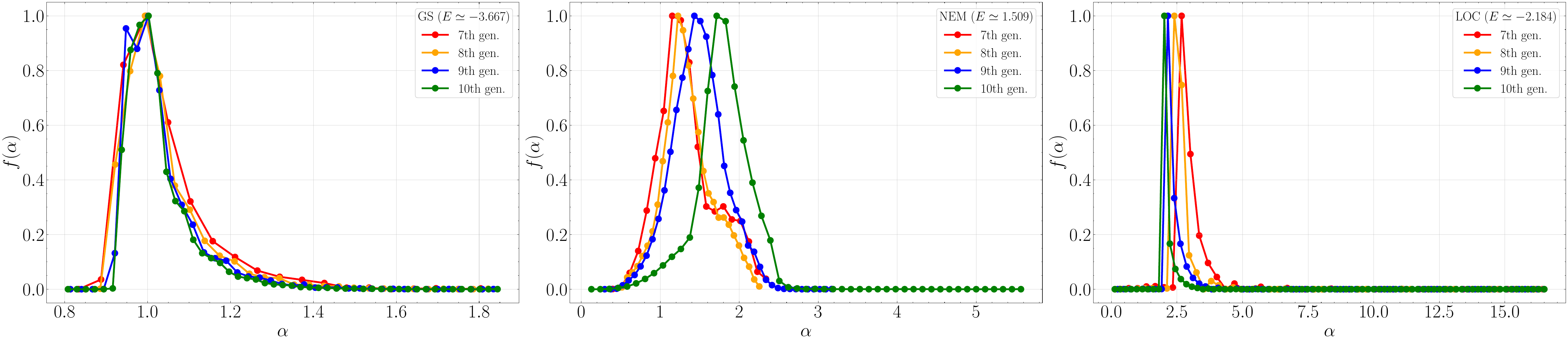}
    \caption{Singularity spectrum $f(\alpha)$ calculated using the direct histogram method for the representative states: Extended (GS, $E \simeq -3.667$, left), Non-Ergodic Multifractal (NEM, $E \simeq 1.509$, middle), and Localized (LOC, $E \simeq -2.184$, right).
    The spectra for generations $g=7$ to $g=10$ provide visual support for our claims, and the distinct character of each state becomes more definite as the system size grows. Although finite, these data provide a reliable basis for the extrapolation methods used in the main text. 
    }
    \label{fig:sm_singspec_hist_canon}
\end{figure*}

\begin{figure}[h!]
    \includegraphics[width=0.7\linewidth]{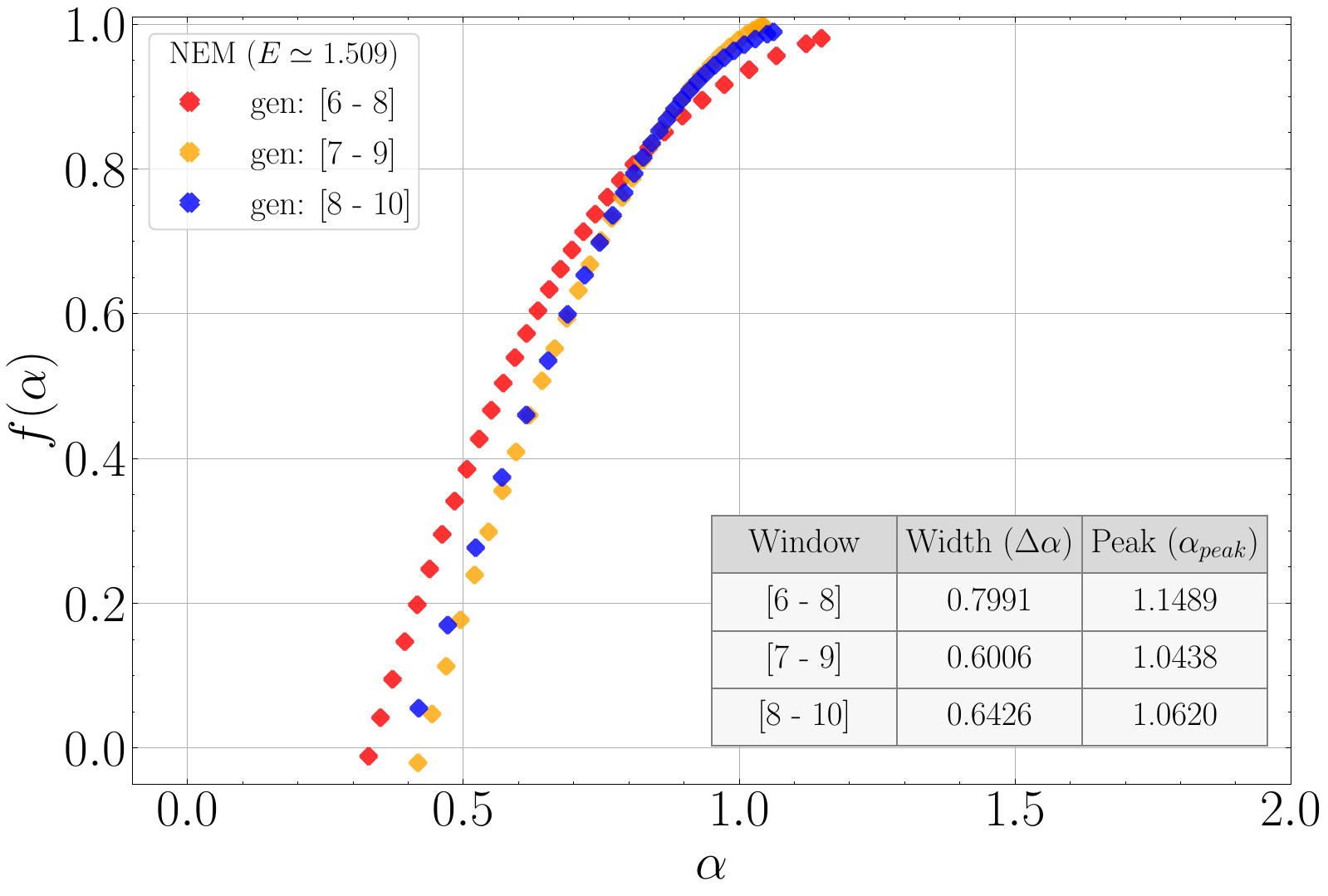}
    \caption{Finite-size scaling of the singularity spectrum $f(\alpha)$ for the NEM state at $E \simeq 1.509$. Each curve represents a sliding window of three consecutive system generations. The inset table lists the corresponding spectrum width ($\Delta\alpha$) and peak position ($\alpha_{\text{peak}}$). Crucially, neither the peak nor the width converges towards the expected ergodic value, a behavior that reinforces the state's non-ergodic nature.}
    \label{fig:sm_sliding_window}
\end{figure}

We present two supplemental analyses of the singularity spectrum. The first, presented in Fig.~\ref{fig:sm_singspec_hist_canon}, is a systematic cross-check using the direct histogram method as defined in the MT (Methods, Eq.~\eqref{eq:singspec_histogram}). The second, in Fig.~\ref{fig:sm_sliding_window}, is a study of the finite-size convergence for the NEM state using a sliding window.

\section{Other examples of NEMs}
\label{sec:sm_other_nems}

\begin{figure}[h!]
    \centering
    \subfigure{%
        \label{fig:tq_therm_nems}
        \includegraphics[width=0.48\linewidth]{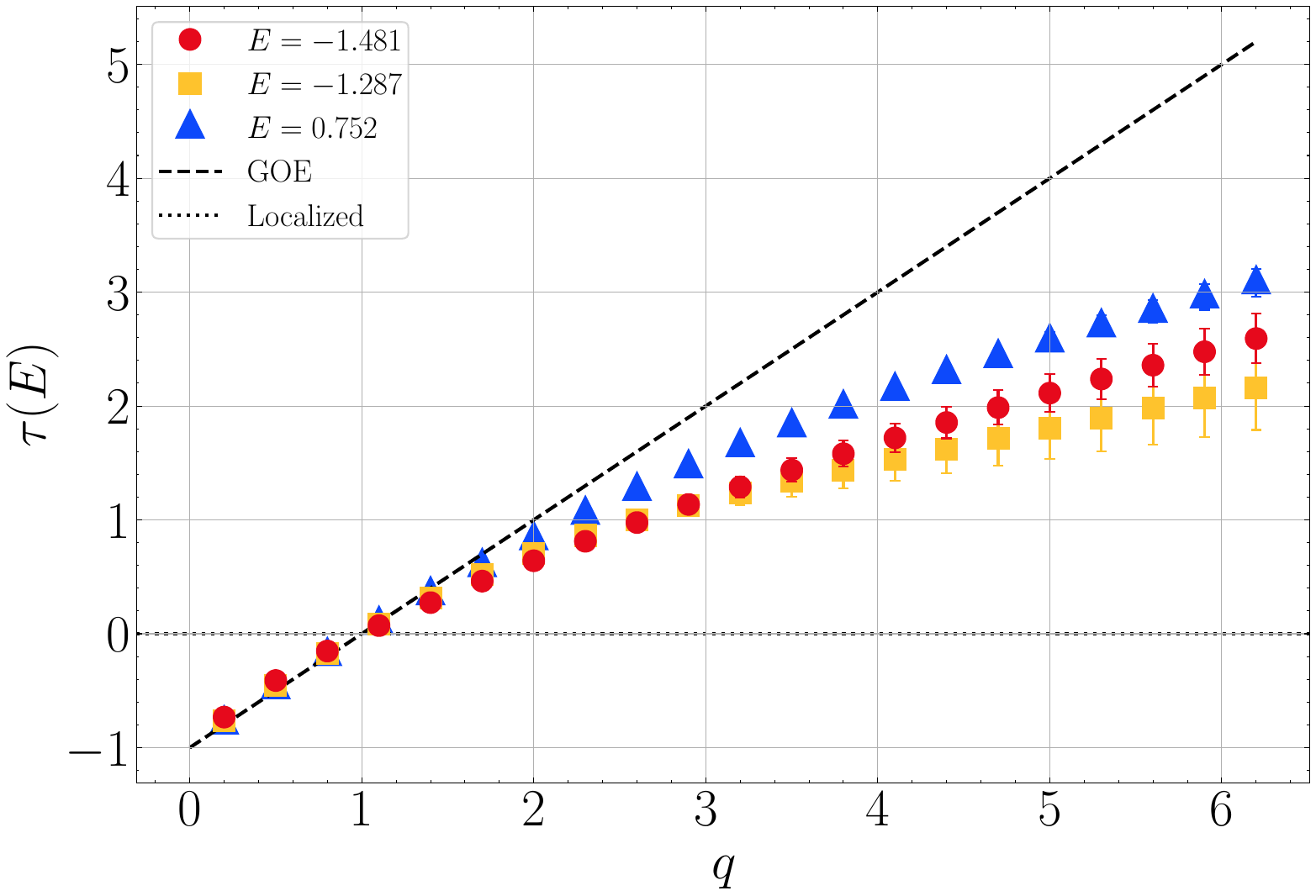}
    }\hspace{\fill}
    \subfigure{%
        \label{fig:gasket_singspec_nems}
        \includegraphics[width=0.48\linewidth]{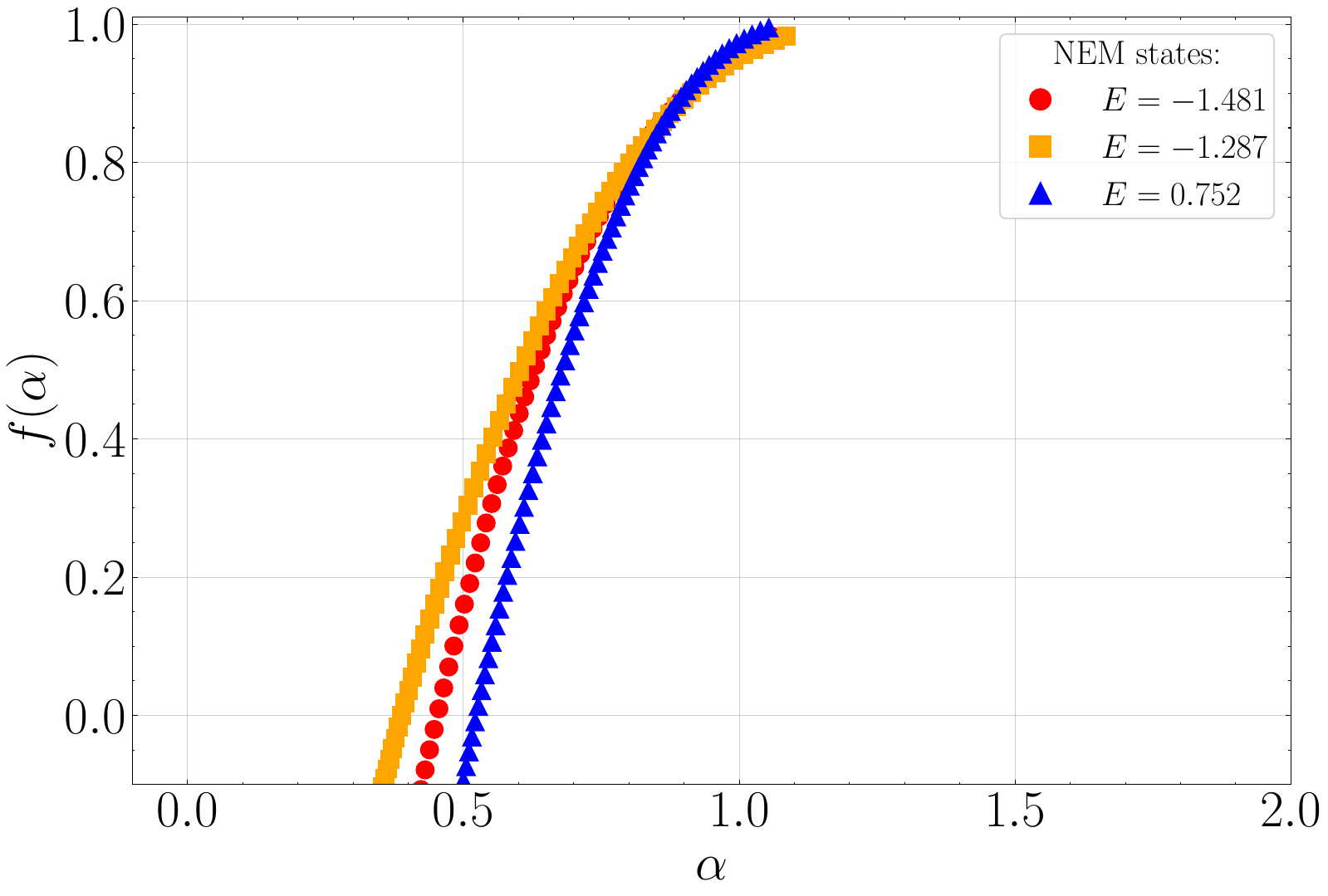}
    }
    \caption{Multifractal analysis of Non-Ergodic Multifractal (NEM) states. 
    Left panel: Thermodynamic mass exponents $\tau(E)$ as a function of the moment order $q$ for three NEM states at energies $E \simeq -1.481$ (red circles), $E \simeq -1.287$ (orange squares), and $E \simeq 0.752$ (blue triangles). The dashed line represents the theoretical expectation for GOE statistics ($D_q=1$, implying $\tau(q) = q-1$), while the dotted line indicates behavior typical of localized states ($\tau(q) \simeq 0$ for $q>0$). 
    Right panel: The singularity spectra $f(\alpha)$ for the same NEM states, illustrating their different strength of multifractality.}
    \label{fig:multifractality_nems}
\end{figure}

\begin{figure}[h!]
    \includegraphics[width=0.75\linewidth]{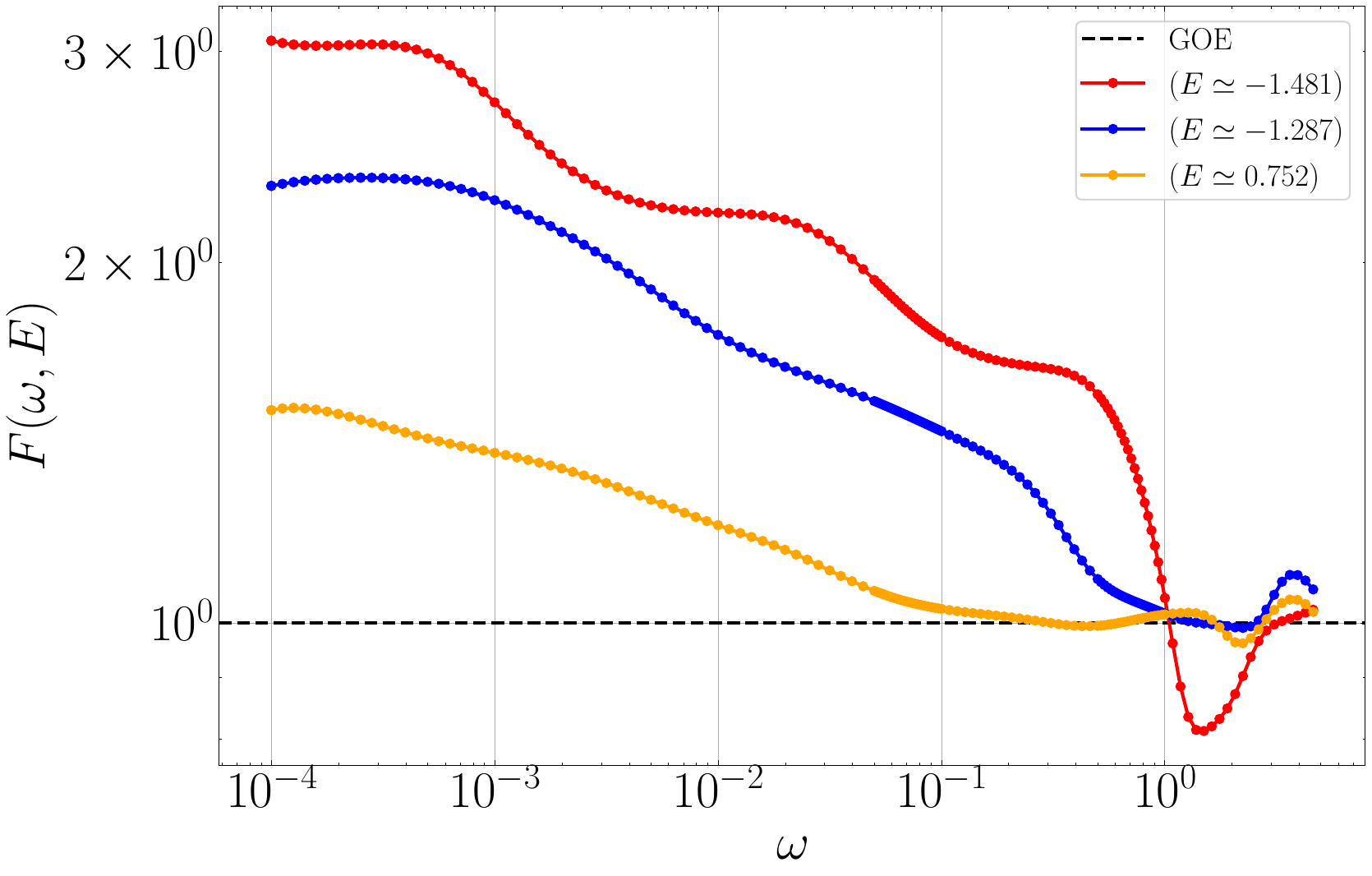}
    \caption{Frequency correlator $\mathcal{F}(\omega, E)$ for three other representative non-ergodic multifractal (NEM) states on a generation $g=6$ SG, compared to the Gaussian Orthogonal Ensemble (GOE) benchmark (dashed line). The states correspond to energies $E \simeq -1.481$ (red), $E \simeq -1.287$ (blue), and $E \simeq 0.752$ (orange). The distinct power-law fall-off of each correlator distinguishes each state.}
    \label{fig:gasket_correlator_nems}
\end{figure}

\begin{figure}[h!]
    \includegraphics[width=\linewidth]{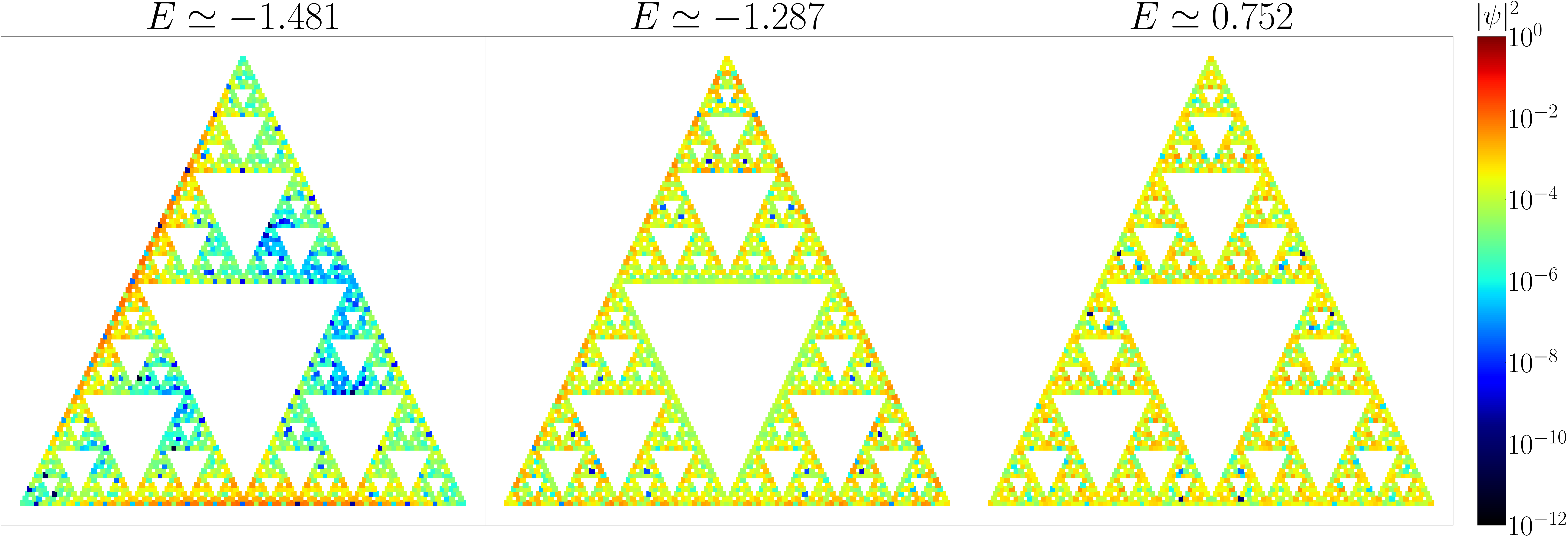}
    \includegraphics[width=\linewidth]{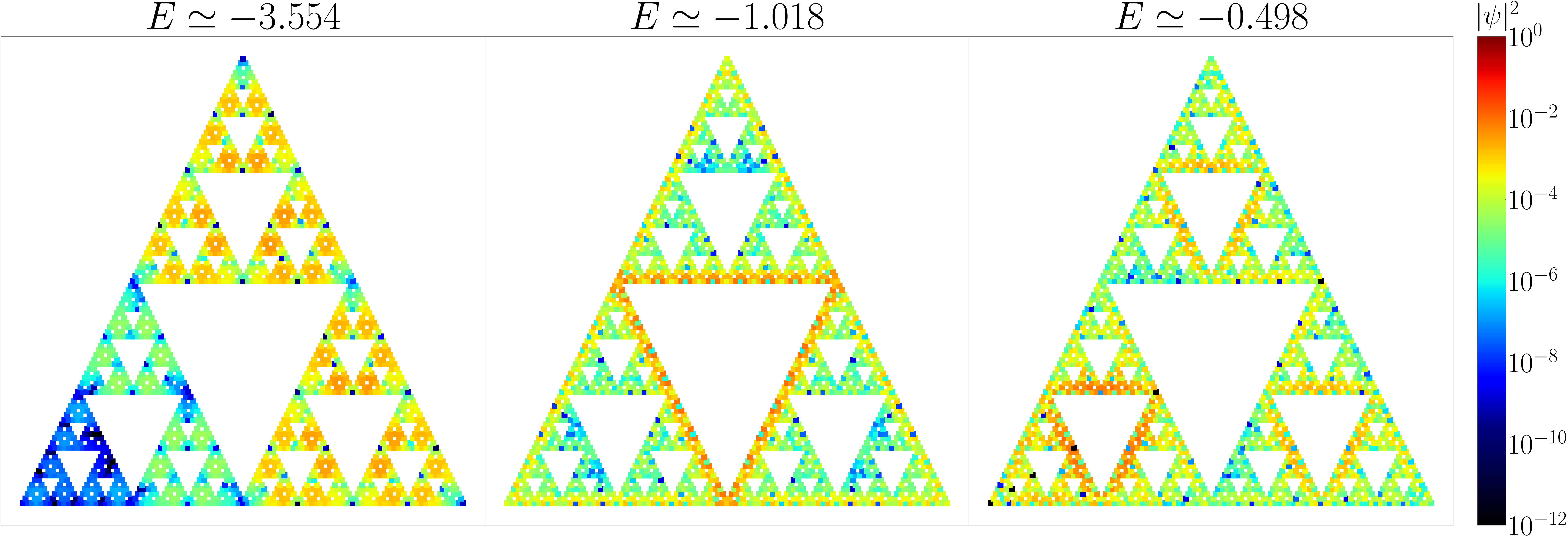} 
    
    \caption{Intensity profiles of different types of Non-Ergodic Multifractal (NEM) states. The color scale indicates $|\psi|^2$ intensity, from negligible (black regions) to high (red regions). \\
    Top row: Intensity distributions for the NEM states detailed in Sec.~\ref{sec:sm_other_nems}. Energies are (from left to right): $E \simeq -1.481$, $E \simeq -1.287$, and $E \simeq 0.752$.
    Bottom row: Other wavefunction structures at energies (from left to right): $E \simeq -3.554$, $E \simeq -1.018$, and $E \simeq -0.498$.
    In these supplementary examples, some states are made with ``pieces of multifractals'' and display properties of either insulators or metals~\cite{corr_cuevas}, while others have spatial features reminiscent of bulk and edge modes previously identified in fractal geometries~\cite{fractals_kempkes, fractals_fischer}.}
    \label{fig:gasket_amplitude_nems}
\end{figure}

In this section, we strengthen our findings on the emergence of NEM states, confirming that they are not numerical anomalies. We present additional data for several representative multifractal states at different energies, all analyzed following the methodologies detailed in the MT. Specifically, their thermodynamic mass exponents $\tau(E)$ and the related singularity spectra $f(\alpha)$ are presented in Fig.~\ref{fig:multifractality_nems}. Furthermore, the spectral density-density correlation functions $\mathcal{F}(\omega, E)$ for these same states are displayed in Fig.~\ref{fig:gasket_correlator_nems}, and their wavefunction intensity profiles are shown in Fig.~\ref{fig:gasket_amplitude_nems}.

\section{Characterization of Weakly Ergodic and Special Multifractal Eigenstates}
\label{sec:charact_multif}

The SG eigenspectrum hosts states with features that go beyond the exemplary cases classified in the MT. This section presents a study of examples of weakly ergodic states \cite{multifractality_khaymovich} (specifically at energies $E \simeq 0.000$ and $E \simeq 2.000$) and also examines the state at $E \simeq 1.000$ that shows special properties. This latter state is remarkable for its robustness and potential for quantum control, as discussed in the MT, warranting a more detailed investigation of its characteristics here. For these states, we employ the same finite-size scaling methodology for mass exponents $\tau_q(E)$ and additionally examine their histogram singularity spectra, wavefunction amplitudes, and statistical properties (such as skewness).

\subsection{Multifractal Analysis}

\begin{figure}[h!]
    \centering
    \subfigure[$E\simeq 0.000$]{%
        \includegraphics[width=0.3\textwidth]{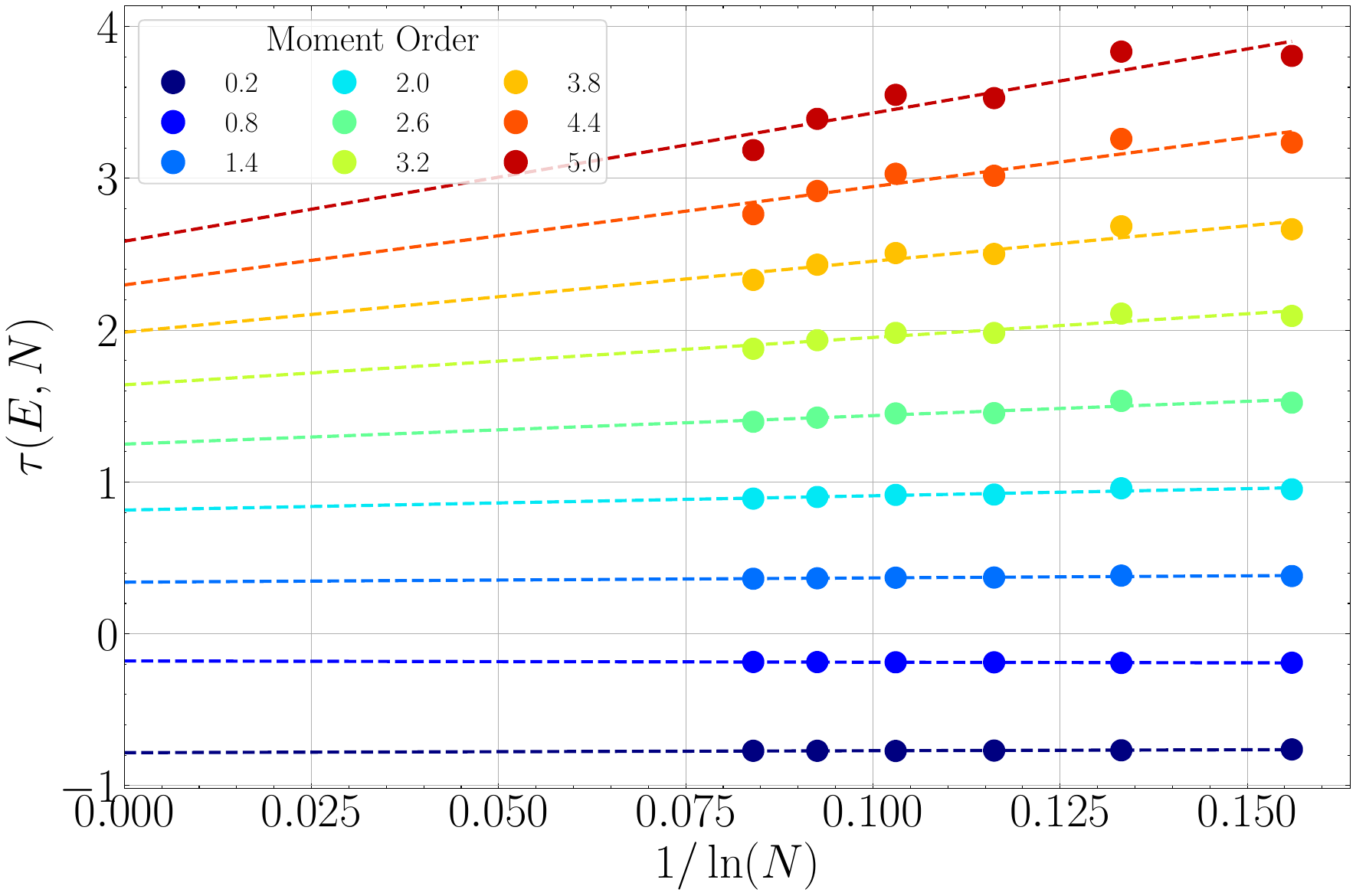}
        \label{fig:tqfinite_E0} 
    }
    \hspace{\fill} 
    \subfigure[$E\simeq 1.000$ ]{%
        \includegraphics[width=0.3\textwidth]{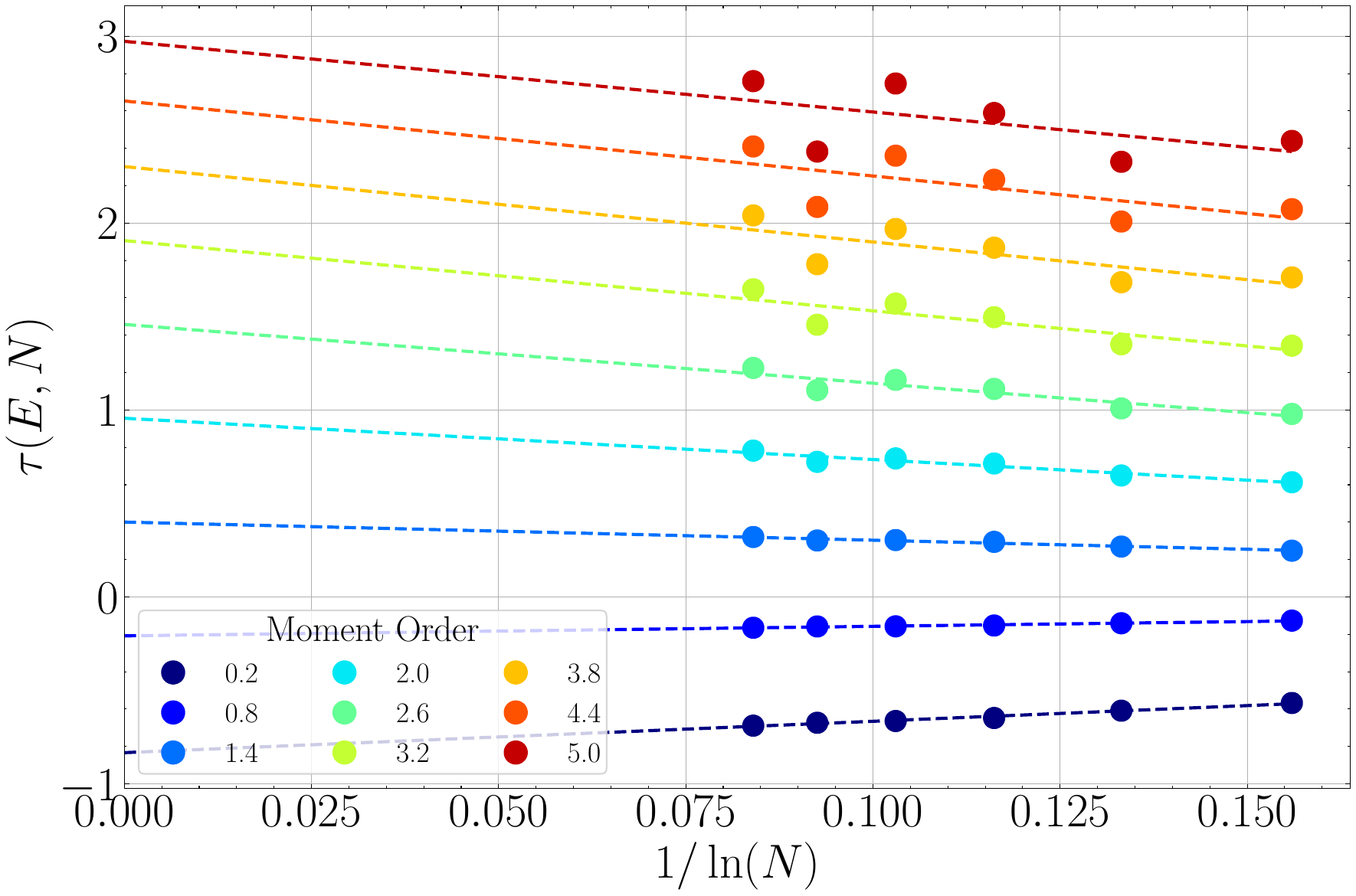}
        \label{fig:tqfinite_E1}
    }
    \hspace{\fill}
    \subfigure[$E\simeq 2.000$]{%
        \includegraphics[width=0.3\textwidth]{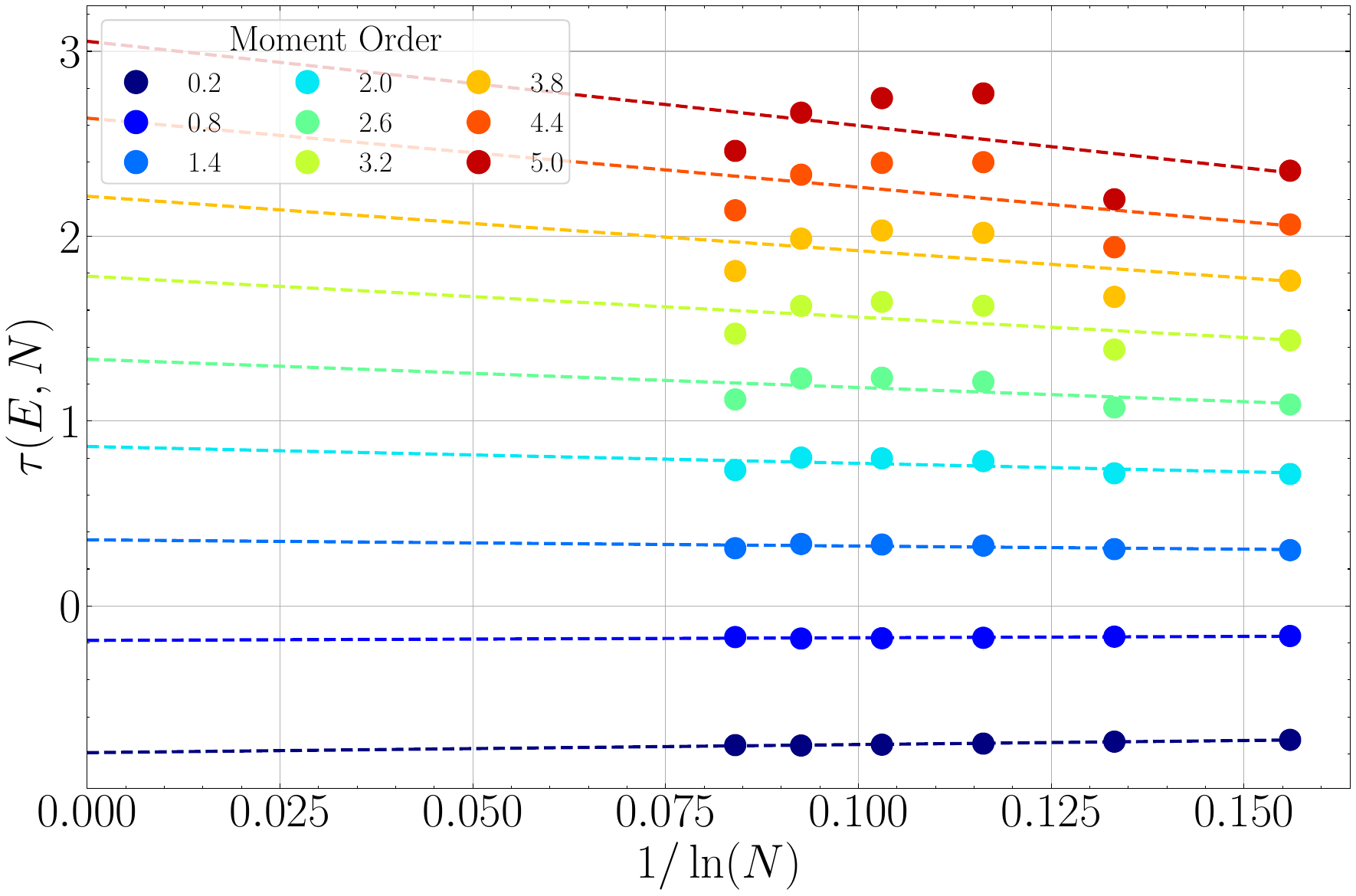}
        \label{fig:tqfinite_E2}
    }
    \caption{Finite-size scaling extrapolation of mass exponents for representative states of energies: $E \simeq 0.000, E \simeq 1.000, E \simeq 2.000$. The thermodynamic value, $\tau_q(E)$, is obtained as the $y$-intercept of a linear fit (dashed lines) to the data points from SG generations $g=5$ through $g=10$.}
    \label{fig:tqfinite_wms}
\end{figure}

\begin{figure*}[h!]
    \centering
    \subfigure{%
        \includegraphics[width=0.49\linewidth]{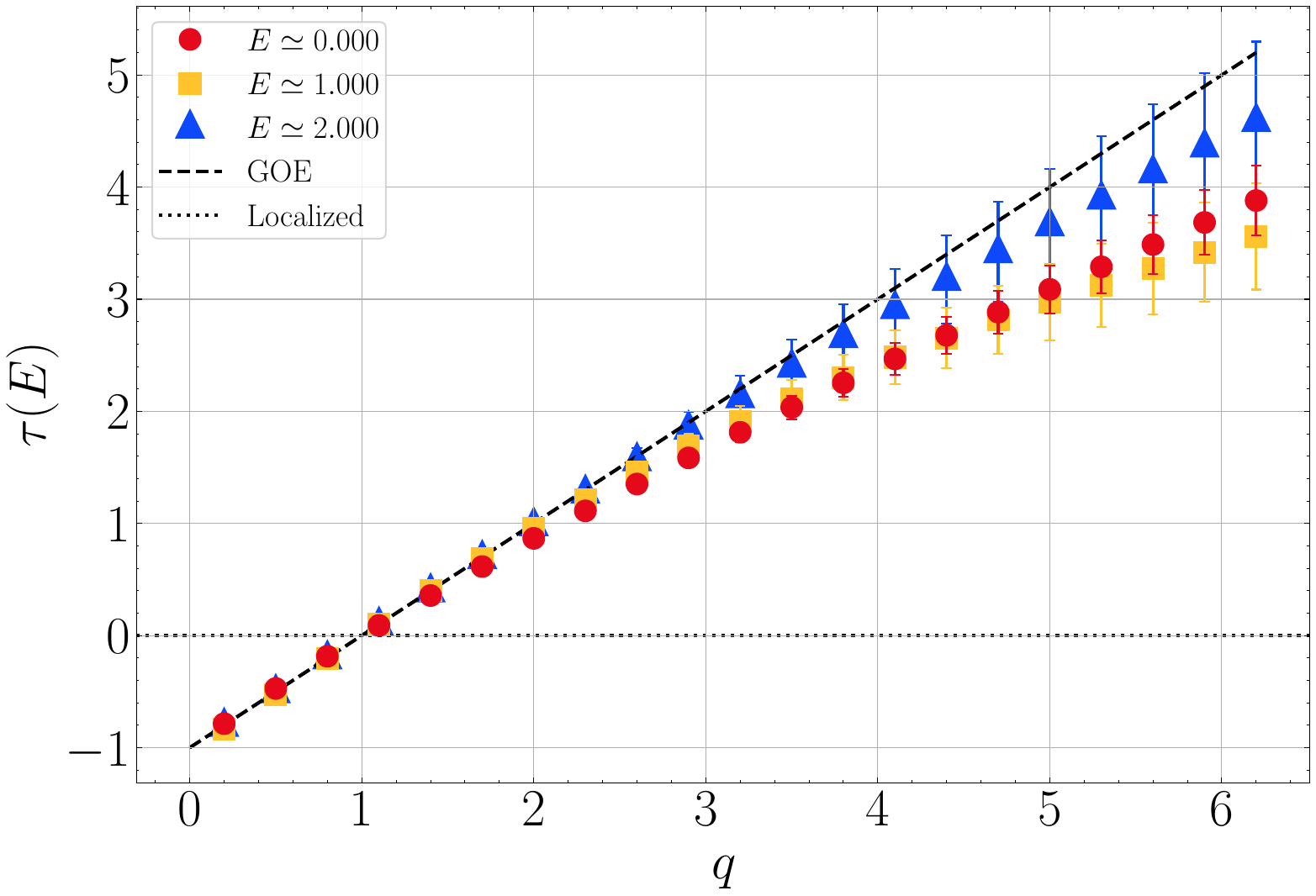}
    }
    \hspace{\fill} 
    \subfigure{%
        \includegraphics[width=0.485\linewidth]{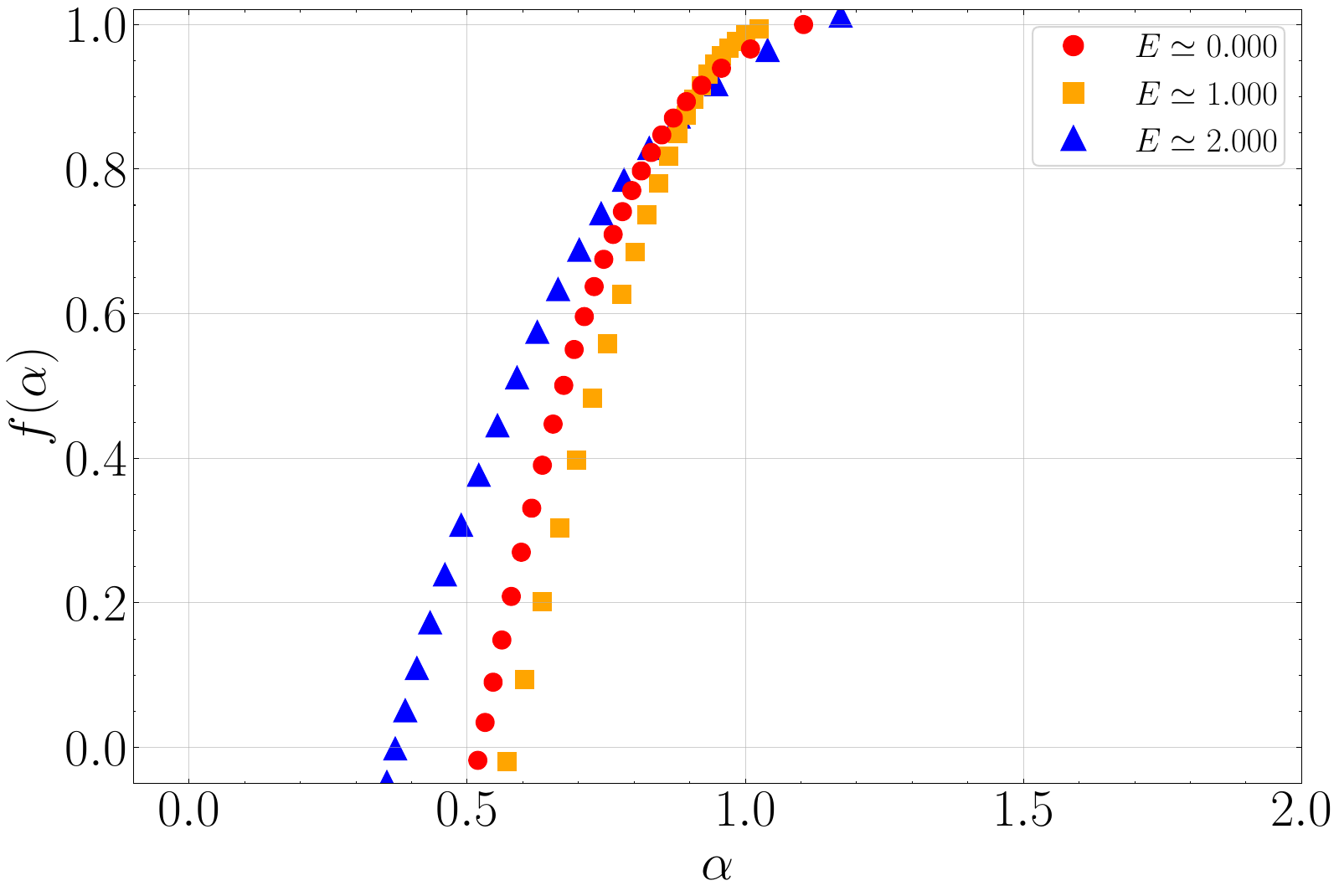}
    }
    \caption{Left panel: Mass exponents $\tau_q(E)$ as a function of moment order $q$for representative states at energies: $E \simeq 0.000$ (red circles), $E \simeq 1.000$ (orange squares), $E \simeq 2.000$ (blue triangles). Statistical errors are included; where not visible, these error bars are smaller than the marker size. The dashed black line indicates the behavior of ideal extended states ($\tau_q = q-1$), while the dotted black line represents ideal localized states ($\tau_q = 0$).\\
    Right panel: The singularity spectra for these states, obtained via Legendre transform, showing a pronounced parabolic shape.}
    \label{fig:multifractality_wms}
\end{figure*}

The finite-size scaling of the effective mass exponents $\tau_q(E, N)$ -- calculated as per Eq.~\eqref{eq:tau_from_ipr} from the MT -- is performed for system sizes ranging from generation $g = 5$ to $g = 10$ and presented in Fig.~\ref{fig:tqfinite_wms}. The thermodynamic values $\tau_q(E)$ are displayed in the left panel of Fig.~\ref{fig:multifractality_wms}.

\begin{figure*}[!htb]
    \includegraphics[width=\textwidth]{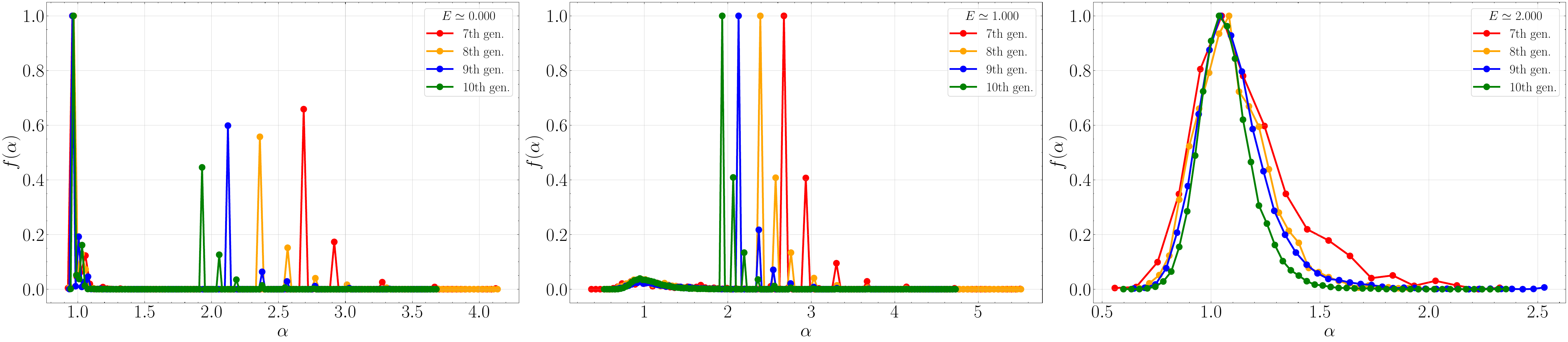}
    \caption{Singularity spectrum $f(\alpha)$ calculated using the direct histogram method for the states at $E \simeq 0.000$ (left), $E \simeq 1.000$ (middle), and $E \simeq 2.000$ (right). Data are shown for generations $g=7$ to $g=10$. The histograms for the state $E \simeq 2.000$ suggest a weakly ergodic character, with the peak near $\alpha \simeq 1$. The states at $E \simeq 0.000$ and $E \simeq 1.000$ display unconventional, comb-like spectra, pointing to their unique spatial distribution across a few disjoint sets of sites.}
    \label{fig:singspec_histogram_wms}
\end{figure*}

Here, one can observe that while for small moment orders $q$ the mass exponents $\tau_q(E)$ for these states tend to follow the $\tau_q \approx q-1$ behavior characteristic of extended systems, deviations emerge at larger $q$. Although the thermodynamic fractal dimension $D_2$ of these states is high ($0.85 \lesssim D_2 \lesssim 0.95$), there are deviations—in particular for $E \simeq 1.000$—in $\tau_q(E)$ at $q>2$, indicating a departure from simple ergodicity and calling for a more detailed examination. The corresponding singularity spectra are displayed in the right panel of Fig.~\ref{fig:multifractality_wms} and in Fig.~\ref{fig:singspec_histogram_wms}.

\subsection{Skewness}

Skewness (Sk) quantifies the asymmetry of a probability distribution \cite{multifractality_roy, multifractal_seckler}. For the wavefunction intensity, $|\psi_E|_i^2$, across lattice sites, $\text{Sk}$ reveals deviations from a symmetric spatial spread. For instance, a perfectly uniform (ideal extended) state would have $\text{Sk}=0$. In contrast, localized states highly concentrated on a few sites with most other sites having near-zero intensity would exhibit large positive skewness. Skewness therefore serves as an additional qualitative indicator of spatial inhomogeneity, complementing the quantitative details of multifractal analysis. It is defined as:

\begin{equation}
    \text{Sk} = \frac{\frac{1}{N}\sum_{i=1}^{N} (|\psi_E|_i^2 - \langle |\psi_E|^2 \rangle)^3}{\left(\frac{1}{N}\sum_{i=1}^{N} (|\psi_E|_i^2 - \langle |\psi_E|^2 \rangle)^2\right)^{3/2}},
\end{equation}

where $|\psi_E|_i^2$ is the wavefunction intensity at site $i$ for an eigenstate $\Phi_E$, $N$ is the total number of active sites for a given system generation $g$ (i.e. $N_g = N$), and the bracket denotes the spatial average. 

\begin{figure*}[h!]
    \includegraphics[width=0.8\linewidth]{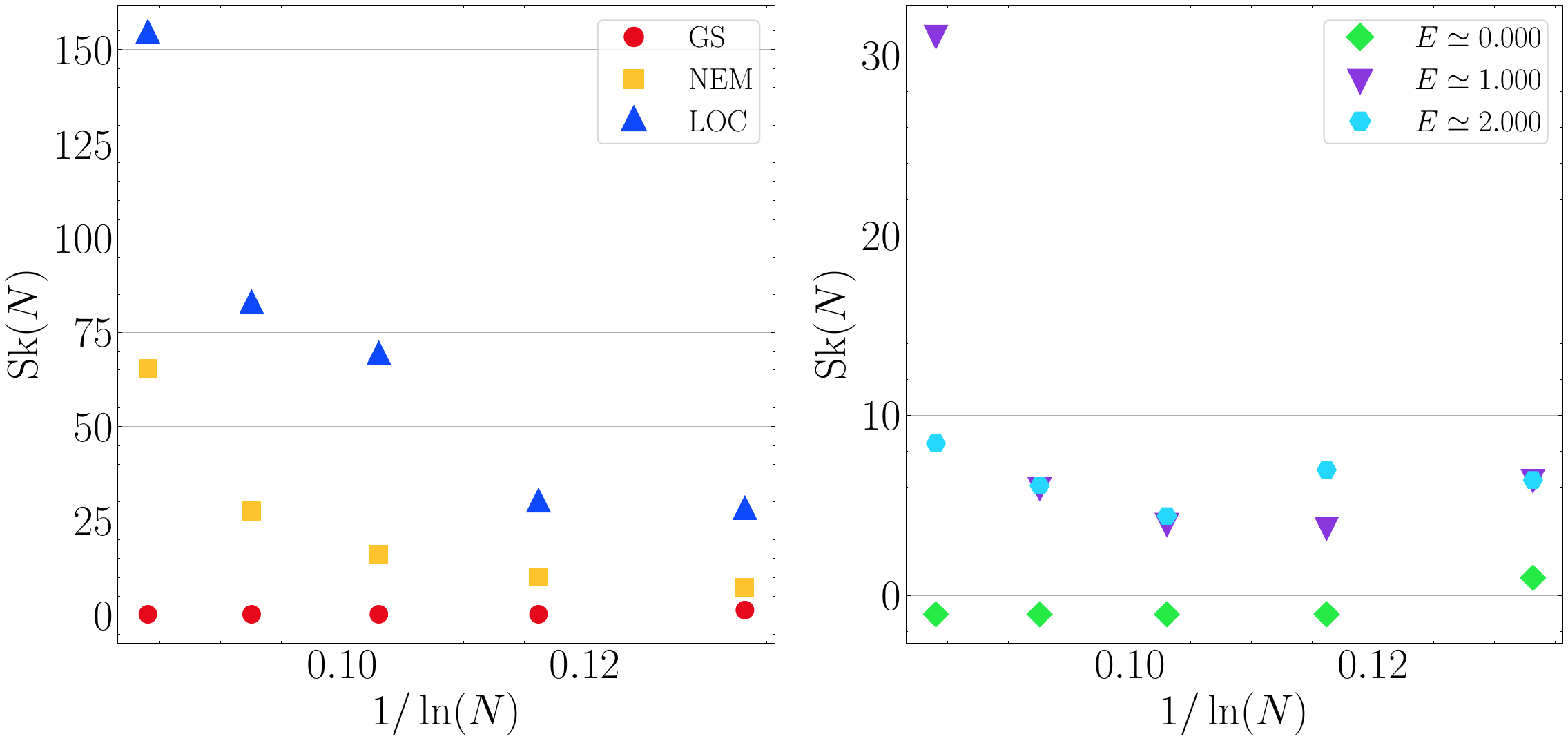}
    \caption{Finite-size scaling of wavefunction intensity skewness $\text{Sk}(N)$ versus $1/\ln N$. 
    (Left panel) Benchmark behaviors for representative extended (GS), multifractal (NEM), and localized (LOC) states. These states are extensively discussed in the Main Text.
    (Right panel) Skewness scaling for eigenstates at energies $E \simeq 0.000$, $E \simeq 1.000$, and $E \simeq 2.000$. 
    Data points correspond to different system sizes and trends indicate the behavior towards the thermodynamic limit.}
    \label{fig:fss_skewness}
\end{figure*}

Fig.~\ref{fig:fss_skewness} presents the finite-size scaling (FSS) of the skewness, $\text{Sk}(N)$, for various characteristic states, obtained by plotting $\text{Sk}(N)$ against $1/\ln N$. The left panel of Fig.~\ref{fig:fss_skewness} establishes benchmark behaviors by displaying the FSS of skewness for the representative extended, localized, and multifractal states discussed in the MT. The extended ground state (GS, red circles) and the localized state (LOC, blue triangles) align with theoretical expectations: the former’s skewness values extrapolate towards $\text{Sk}_\infty \approx 0$ (consistent with its relatively uniform spatial spread), whereas the latter has a large, positive skewness that increases with system size $N$ (as for strong concentration on few sites). The representative non-ergodic multifractal state (NEM) also displays significant positive skewness for finite system sizes, trending towards a smaller yet distinctly positive $\text{Sk}_\infty$. Our analysis primarily uses this observed trend to differentiate the NEM's behavior from both the near-zero skewness characteristic of extended states and the large skewness found for localized states. With this in mind, in the right panel of Fig.~\ref{fig:fss_skewness} we present the skewness FSS for the eigenstates at $E \simeq 0.000$, $E \simeq 1.000$, and $E \simeq 2.000$. 

\begin{figure*}[h!]
    \includegraphics[width=1\linewidth]{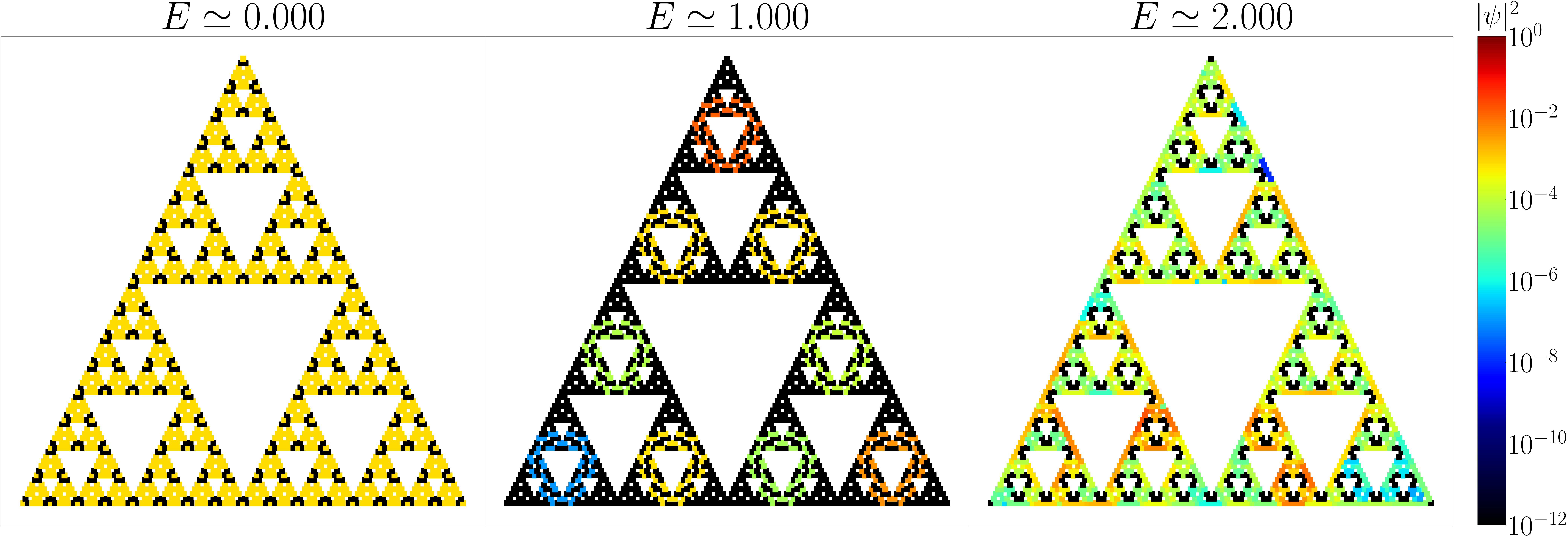}
    \caption{Intensity profiles of $|\psi|^2$ corresponding states at energies (from left to right): $E \simeq 0.000$, $E \simeq 1.000$, and $E \simeq 2.000$. The states for $E \simeq 0.000$ and $E \simeq 2.000$ are weakly ergodic. The state for $E \simeq 1.000$ is a case of clustered-patterned multifractality. The color intensity, indicated by the colorbar, represents $|\psi|^2$ on a logarithmic scale.}
    \label{fig:gasket_amplitude_msc}
\end{figure*}

The state at $E \simeq 0.000$ (green diamonds) is notable as it converges to a distinct, negative skewness, $\text{Sk}_\infty \simeq -1.055$ (determined from individual FSS fits not shown in this figure). Negative skewness typically implies that the distribution has a long tail towards lower intensity values: the majority of participating sites have higher-than-average intensities, with a sharp drop-off at the very low-intensity end. The amplitude profile of this state (Fig.~\ref{fig:gasket_amplitude_msc}a) visually corroborates this interpretation of its asymmetry; this observation, when considered alongside the mass exponent characterization, supports its classification as weakly ergodic.

The eigenstate at $E \simeq 2.000$ (cyan hexagons) is also classified as weakly ergodic. Although its skewness is relatively stable and extrapolates to a small but positive $\text{Sk}_\infty$, its amplitude profile (Fig.~\ref{fig:gasket_amplitude_msc}c) reveals a broad, non-uniform distribution of intensity across many sites, with sharp probability decays occurring only at the edges of the larger groups of voids of the SG. These features, combined with the state's high mass exponent value, point towards its classification also as weakly ergodic. 

Finally, the state at $E \simeq 1.000$ (purple triangles) exhibits a clear positive extrapolated skewness. From the FSS, its $\text{Sk}(N)$ values drift towards a moderate positive value for larger system sizes. This trend, while distinct from those of the weakly ergodic states at $E \simeq 0.000$ and $E \simeq 2.000$, instead appears more similar to the benchmark NEM. Its probability distribution, while not uniform, is situated on a structured subset of sites with significant intensity fluctuations, leading to its positive skewness from extended regions of low probability. This aligns with its intensity profile (Fig.~\ref{fig:gasket_amplitude_msc}b), which reveals clusters of varying intensities around the $g-3$ sub-gasket blocks, supporting its characterization as a form of 'clustered-patterned multifractality'.

\section{Robustness}
\label{sec:sm_robustness}

\subsection{Local Robustness of Multifractality}

\begin{figure}[h!]
    \centering
    \subfigure{%
        \includegraphics[width=0.48\linewidth]{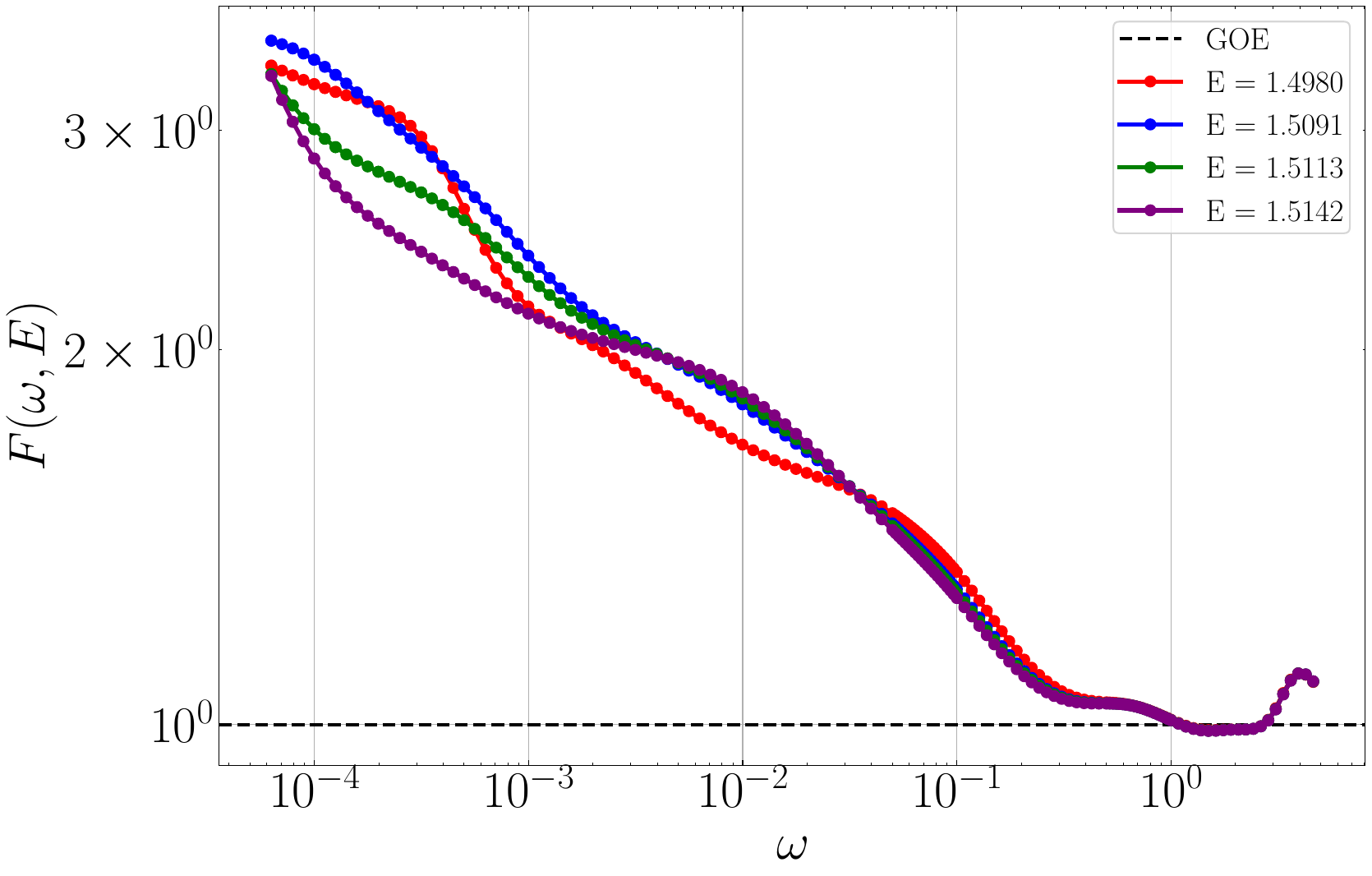}
    }\hspace{\fill}
    \subfigure{%
        \includegraphics[width=0.48\linewidth]{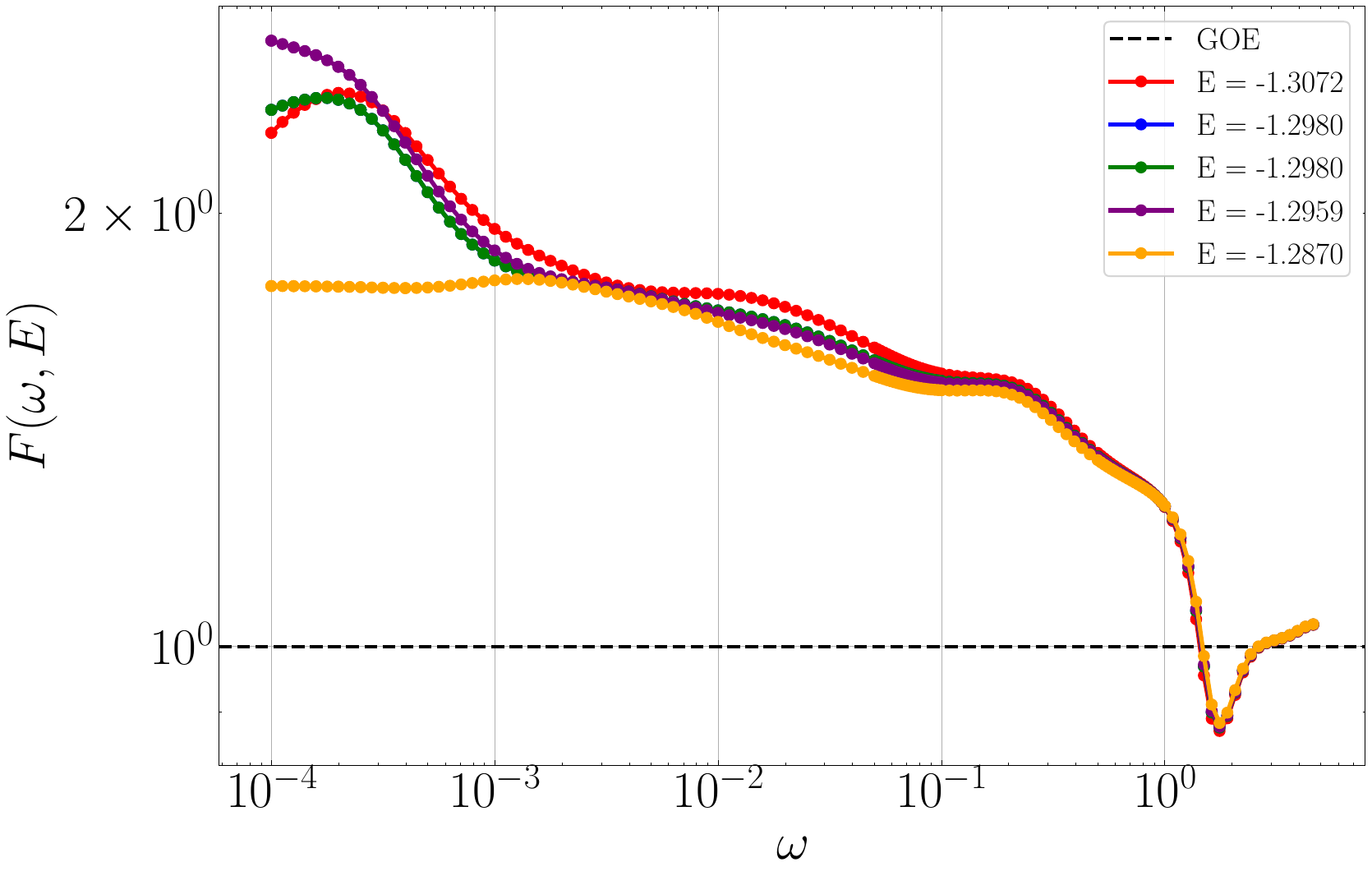}
    }
    \caption{
    Stability of the two-point correlator $\mathcal{F}(\omega, E)$ for states within a finite energy window. 
    Analysis for generation $g=6$ states in the vicinity of the representative NEM at $E \simeq 1.509$ (left) and around $E \simeq -1.297$ (right). The persistence of the general power-law decay points to a robust multifractal character, though quantitative differences in plateau heights and decay exponents are apparent.
    The dashed line indicates the Gaussian Orthogonal Ensemble (GOE) value for ergodic states.
    }
    \label{fig:corrf2_robust}
\end{figure}

To test whether the revealed NEMs are isolated states---requiring infinite precision to detect---we analyze the stability of the two-point correlator $\mathcal{F}(\omega, E)$ against small variations in energy. The results, shown in Fig.~\ref{fig:corrf2_robust} for two spectral regions, reveal that while the overall power-law behavior is preserved across the energy neighborhood, quantitative details may vary between nearby states. This observation indicates that NEMs are robust spectral features forming a mini-band~\cite{singspec_kravtsov}, rather than isolated occurrences.

\subsection{Robustness Against Randomly Displaced Defects}

\begin{figure}[!h]
    \includegraphics[scale=0.5]{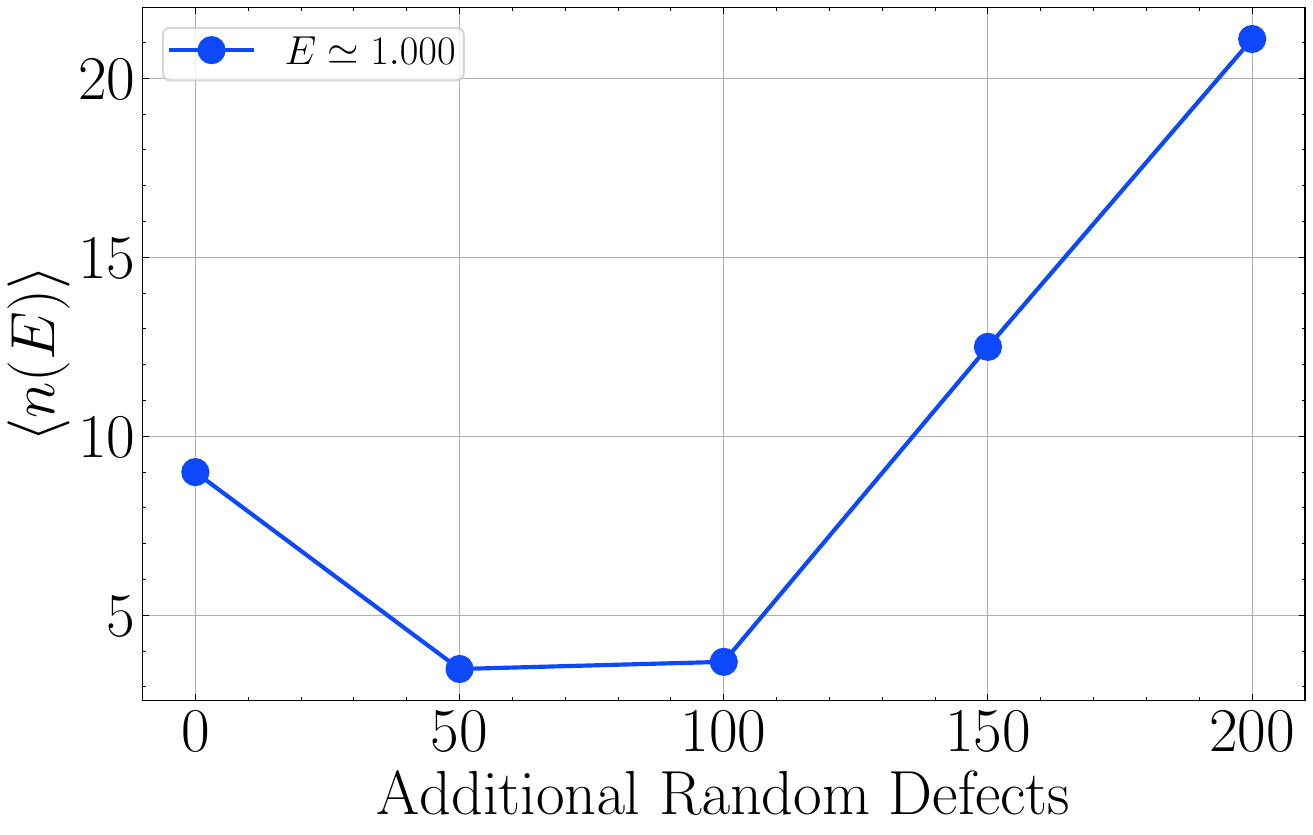}
    \caption{Change in the average degeneracy $\langle n(E) \rangle$ of the multifractal state at $E \simeq 1.000$ in a $g=6$ SG as additional random defects are introduced. The plotted values are averages over $p=10$ distinct disorder realizations.}
    \label{fig:mcs_ardcount}
\end{figure}

\begin{figure}[!htb]
    \includegraphics[width=1\linewidth]{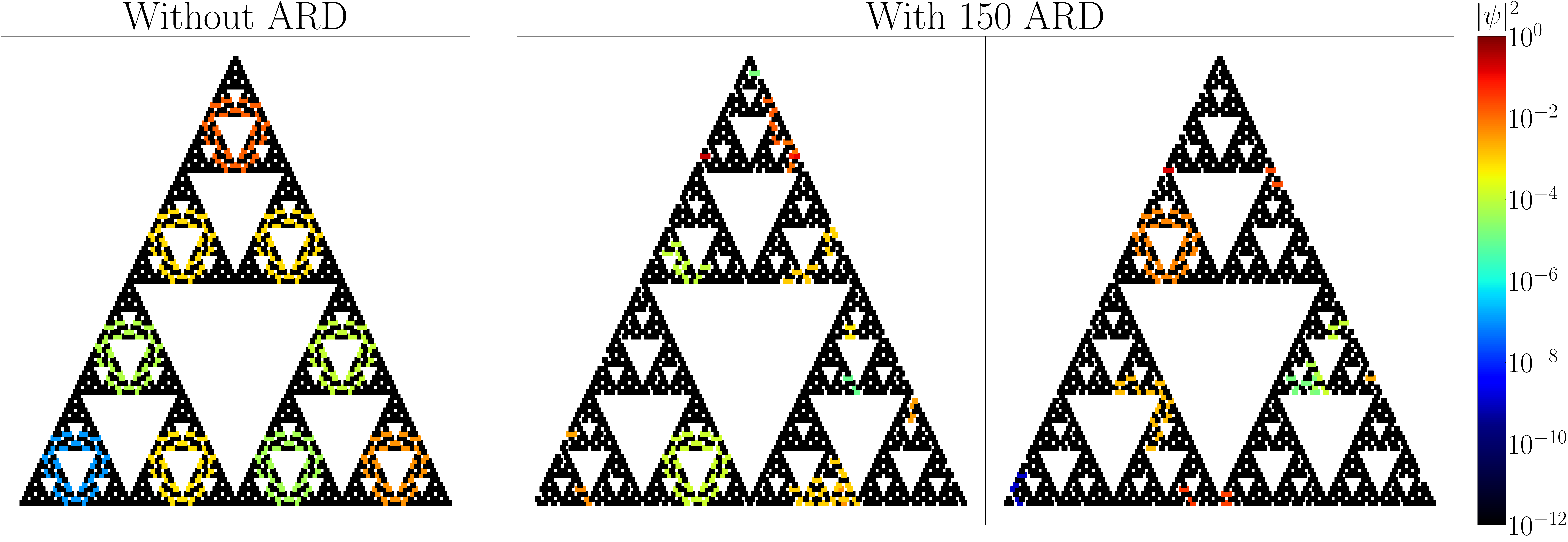}
    \caption{Intensity profiles of the multifractal state at energy $E \simeq 1.000$ on a $g=6$ SG. Left panel: The state in the pristine system, without additional random defects (ARDs). Center and right panels: Two distinct realizations of the state when 150 ARDs are introduced.
    While some of the clustered features may persist, the introduction of ARDs can generate new quantum interference patterns, and the intensity is redistributed primarily towards the lower hierarchical levels of the SG structure.}
    \label{fig:mcs_with_ard}
\end{figure}

Fig.~\ref{fig:mcs_ardcount} illustrates the evolution of the degeneracy for the special multifractal state at $E \simeq 1.000$ as a function of the number of randomly introduced defects. Contrary to the common expectation that adding defects would disrupt the fractal's self-similar structure and lift the degeneracy of this energy level, an opposite trend is observed. The introduction of random defects may enhance certain quantum interference patterns, leading to an increased degeneracy. The changes to its spatial intensity profile are displayed in Fig.~\ref{fig:mcs_with_ard}, which compares representative configurations before and after the addition of these random defects.

\subsection{Next-Nearest-Neighbor Hopping}
\label{sec:sm_nnn_hopping}

In this section, we investigate the robustness of the emergent multifractality against the inclusion of next-nearest-neighbor (NNN) hopping terms. The modified Hamiltonian, extending the model defined in Eq.~\eqref{eq:hamiltonian} from the MT with NNN interactions, is given by:

\begin{equation}
    \label{eq:hamiltonian_NNN}
    H = \sum\limits_{\substack{\langle i,j \rangle \\ i, j \notin \mathcal{B}}} t\, c^\dagger_i c_j + \sum_{k \in \mathcal{B}} V\, c^\dagger_k c_k + \sum_{\substack{\langle i,k \rangle \\  k \in \mathcal{B} }} \overline{t}\, c^\dagger_i c_k + \sum\limits_{\substack{\langle\langle i,l \rangle\rangle \\ i, l \notin \mathcal{B}}} t'\, c^\dagger_i c_l + h.c.
\end{equation}

Here, $t'$ represents the NNN hopping strength, with double brackets denoting summation over the next-nearest neighbors. For this analysis, parameters other than $t'$ are maintained consistently with those in the MT: $t = 1$, $\overline{t} = 10^{-9}$, and $V = 10^4$. This investigation consists of a qualitative study of the two-point density-density correlator $\mathcal{F}(\omega, E)$ (as defined in Eq.~\eqref{eq:2eig_corr} of the MT), where we probe its behavior across different NNN hopping strengths $t'$.

\begin{figure}[!htb]
    \includegraphics[width=1\linewidth]{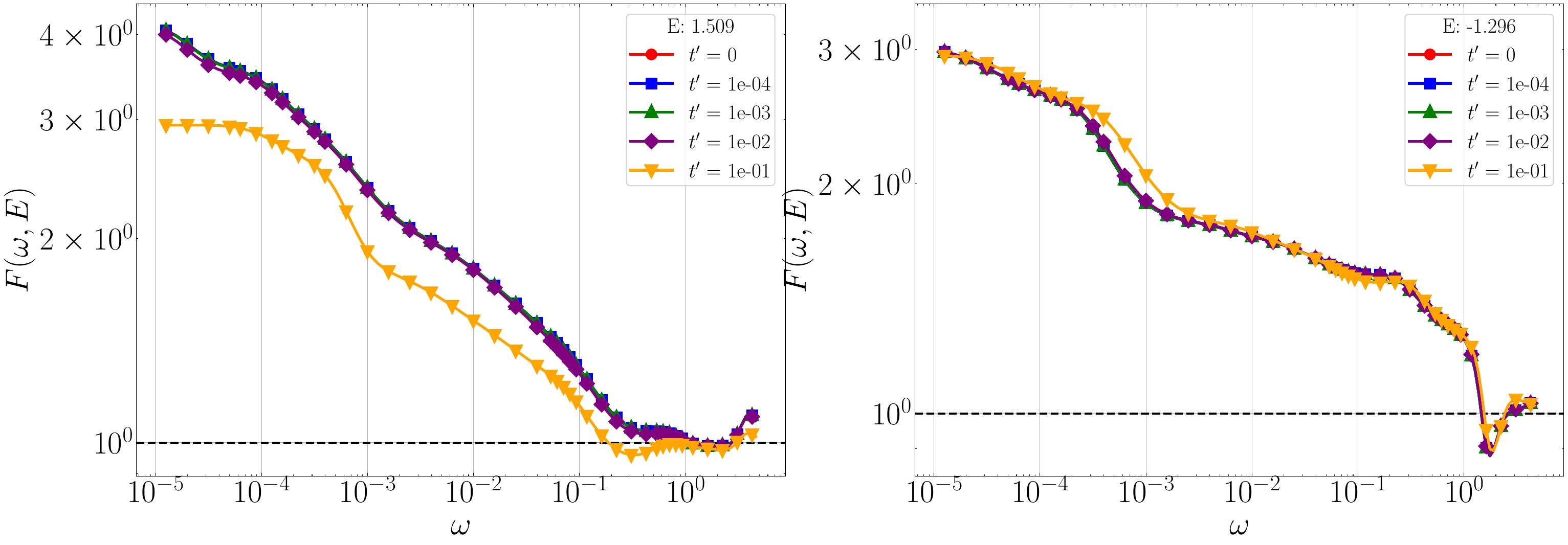}
    \caption{The frequency correlator $\mathcal{F}(\omega, E)$ for representative non-ergodic multifractal (NEM) states, illustrating the impact of varying next-nearest-neighbor (NNN) hopping strengths $t'$. The plots reveal how characteristics such as plateau width and slope behavior respond to changes in $t'$. Lines serve as guides to the eye.}
    \label{fig:corrf2_NNN}
\end{figure}

Fig.~\ref{fig:corrf2_NNN} presents the behavior of $\mathcal{F}(\omega, E)$ for NEM states, specifically at energies $E \simeq 1.509$ and $E \simeq -1.287$, computed on a $g=6$ SG. The introduction of a small NNN hopping strength, up to $t' = 10^{-2}$, does not significantly modify the key qualitative features of the correlator, such as its plateau height and characteristic power-law decay exponent.

However, when the NNN hopping strength is increased to $t' = 10^{-1}$, discernible changes become apparent. The response to this stronger coupling varies among the group of states considered: for some, $t' = 10^{-1}$ induces effects such as a lowering of the correlator's plateau or an alteration of its power-law decay characteristics. Such modifications may signal a shift towards a different multifractal regime or, in particular instances, hint at a more profound transformation of the state's structure, potentially even suppressing its initial multifractal traits. For other states within the group, the impact of NNN hopping remains less pronounced, even at this increased magnitude. Collectively, these findings suggest a degree of robustness in the NEM characteristics against weak NNN perturbations, while stronger NNN coupling can more substantially influence their finer details.

\section{Generalization to Other Fractal Geometries and Interactions}

An intriguing question arising from our findings is the extent to which the emergence of NEMs is specific to the SG, or if it represents a more general phenomenon. From our results, we believe NEMs can manifest in diverse fractal structures, though their characteristics would likely depend on intrinsic geometric properties such as the ramification number—a factor known to be significant in related studies on general graphs and fractals \cite{spectral_dasilva, fractals_ivaki}. 

In this spirit, investigating fractals embedded in three-dimensional space may offer a concrete pathway for observing NEM signatures \cite{sizescaling_rodriguez}. However, performing comprehensive numerical analyses within three-dimensional configurations presents significant computational challenges, currently limiting extensive quantitative verification.

Similarly, the nature of hopping interactions beyond nearest-neighbor or next-nearest-neighbor terms, such as the inclusion of specific long-range hopping models, offers another rich avenue for future theoretical exploration. These broader considerations, while beyond the scope of the current work, may reveal additional mechanisms to enhance or tune multifractality in such structures.

As a step towards exploring different geometries, and to consider a system with potentially more accessible experimental realization, we investigated the negative Vicsek fractal, as detailed below.

\subsection{Negative Vicsek Fractal}
\label{sec:sm_vicsek}
\begin{figure}[!htb]
    \centering
    \includegraphics[width=0.6\linewidth]{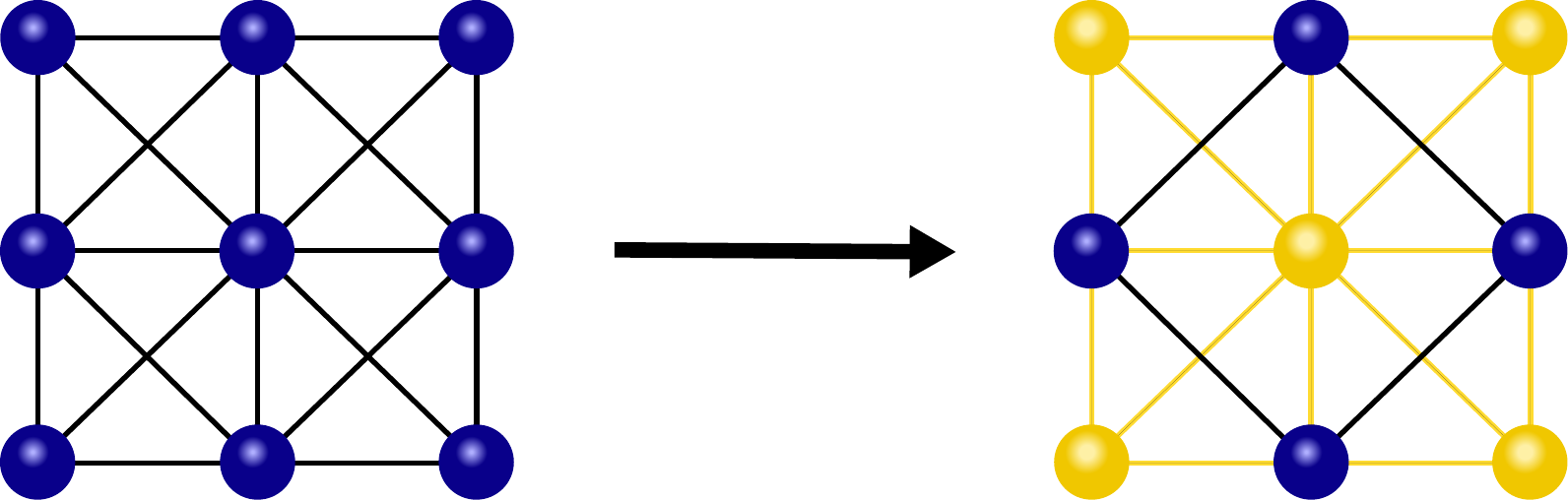}
        \caption{Transformation of a minimal square lattice into the first-generation negative Vicsek fractal. Next-nearest-neighbor hopping is required to keep the lattice connected.}
    \label{fig:zoom_square_to_vicsek}
\end{figure}

\begin{figure}[!htb]
    \centering
    \includegraphics[width=0.8\linewidth]{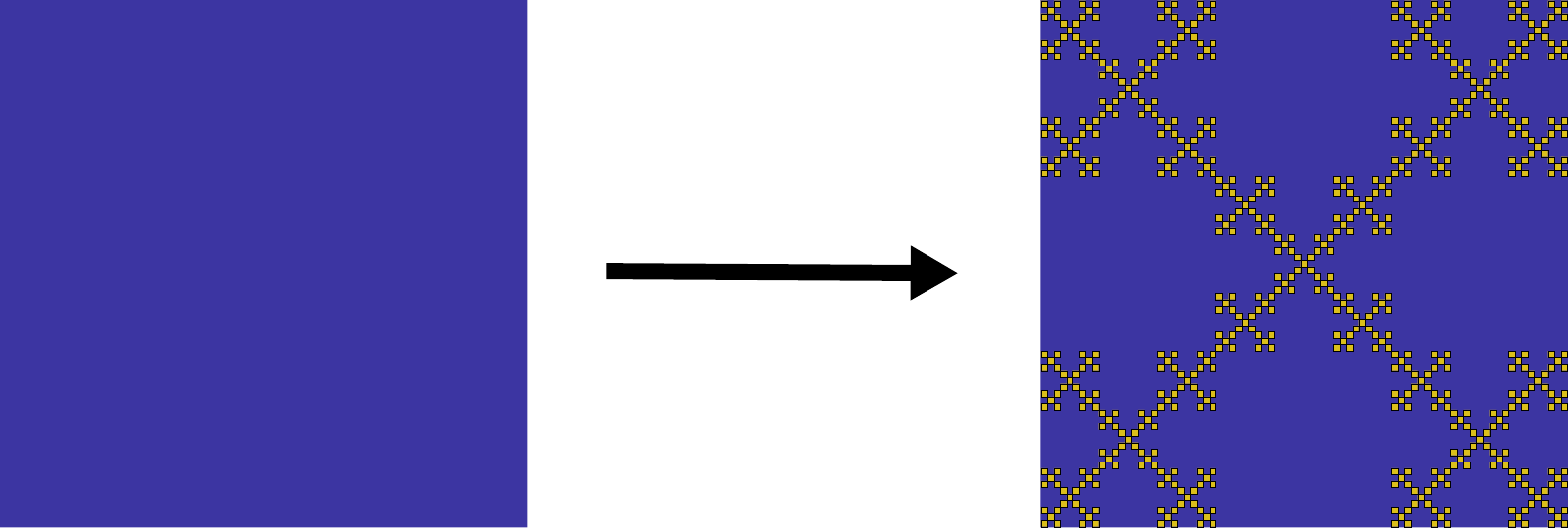}
     \caption{Transformation of a square lattice into the fourth-generation negative Vicsek fractal. The final structure emerges from successive iterations of the elementary transformation, where lattice sites in subset $\mathcal{B}$ are progressively replaced by defects. The resulting fractal is characterized by patching large regions of intact lattice without defects and regions with intricate defect patterns.}
    \label{fig:square_to_vicsek}
\end{figure}

Conceptually, the negative Vicsek (NV) fractal is easier to fabricate than the SG, as it requires significantly fewer defects to be placed in a controllable manner. The starting point is a uniform square lattice $\mathcal{S}$ of $N$ sites. Then, unlike in the SG case, we retain sites corresponding to voids of the fractal and remove all other sites, see Figs.~\ref{fig:zoom_square_to_vicsek} and ~\ref{fig:square_to_vicsek}. We denote the set of indices of the removed sites as $\mathcal{B} \subset \mathcal{S}$.

As in the SG case, each defect at position $r_i$ is modeled by a delta-peaked potential, $V_i = V\,\delta(r - r_i)$, with a strength $V$ much larger than all other energy scales in the system. However, in the NV model, we also account for next-nearest-neighbor interactions, denoted by $t'$. These additional couplings are fundamental, as they maintain connectivity between otherwise isolated regions of the fractal. Defect couplings are weak. The corresponding Hamiltonian reads:
    \begin{equation}
        H = \sum\limits_{\substack{\langle i,j \rangle \\ i, j \notin \mathcal{B}}} t\, c^\dagger_i c_j +  \sum\limits_{\substack{\langle \langle i,k \rangle \rangle \\ i, k \notin \mathcal{B}}} t'\, c^\dagger_i c_k + \sum_{l \in \mathcal{B}} V\, c^\dagger_l c_l + \sum_{\substack{\langle j,l \rangle \\  l \in \mathcal{B} }} \overline{t}\, c^\dagger_j c_l + \sum_{\substack{\langle\langle k,l \rangle \rangle \\  l \in \mathcal{B} }} \overline{t'}\, c^\dagger_k c_l + \text{h.c.}
    \label{vicsek_hamiltonian}
    \end{equation}

Here, single brackets denote summations over the nearest neighbors, and the double brackets over the next-nearest neighbors. 

\subsection{$\text{NPR}_2$ of States for the Negative Vicsek Lattice}

\begin{figure}[!htb]
    \centering
    \includegraphics[width=0.75\linewidth]{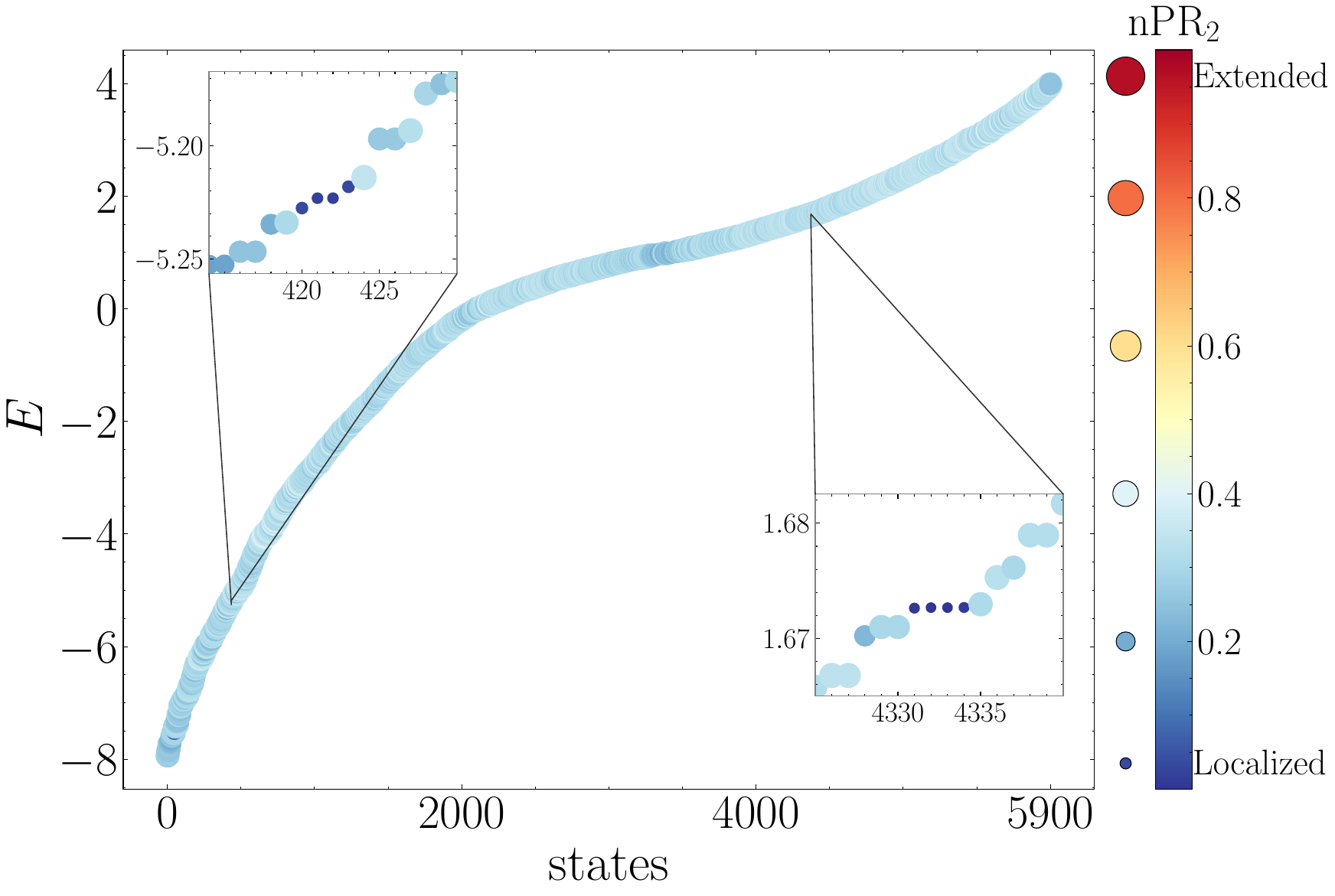}
    \caption{Energy spectrum of the single-particle tight-binding model on the fourth-generation negative Vicsek fractal. The colorbar represents the normalized Participation Ratio ($\text{nPR}_2$) for each eigenstate: small blue markers signify localized states (low $\text{nPR}_2$), while large red markers indicate ergodic/extended states (high $\text{nPR}_2$). Although the large "clean" regions within the NV fractal structure might be expected to suppress the formation of NEMs, the fractal's underlying geometric complexity still holds potential for the emergence of NEMs in certain spectral regions.}
    \label{fig:vicsek_spectrum}
\end{figure}

In Fig.~\ref{fig:vicsek_spectrum}, we present the $\text{nPR}_2$ values for all eigenstates in the energy spectrum of the fourth-generation NV fractal. The complete set of eigenstates and eigenvalues was obtained by exact diagonalization of the Hamiltonian in Eq.~\eqref{vicsek_hamiltonian} with open boundary conditions. The parameters are set as $t = t' = 1$, $ \bar{t} = \bar{t'} = 10^{-9}$, and $V = 10^4$.

